\documentclass[aps, prx, floats, floatfix, twocolumn, nofootinbib, amssymb,
               longbibliography, superscriptaddress]{revtex4-1}

\usepackage{graphicx, bm, hyperref}
\usepackage[usenames]{color} 
\usepackage[normalem]{ulem} 
 
\providecommand{\BibitemShut}[1]{} 

\newcommand{\third}{{\textstyle\frac{1}{3}}}
\newcommand{\half}{{\textstyle\frac{1}{2}}}

\newcommand{\sixth}{{\textstyle\frac{1}{6}}}

\newcommand{\rmd}{\mathrm{d}}
\newcommand{\gaugeD}{\mathcal{D}}
\newcommand{\scri}{\mathcal{I}^+\,}
\newcommand{\four}{\,{}^{(4)}}

\newcommand{\two}{\,{}^{(2)}}
\newcommand{\tN}{\tilde N}
\newcommand{\tphi}{\tilde{\phi}}
\newcommand{\tr}{\mathrm{tr} \,}
\newcommand{\diff}{\mathrm{d}}
\newcommand{\const}{\mathrm{const}}
\newcommand{\hateq}{\; \hat{=} \;}
\newcommand{\etal}{{\it et al.\ }}

\usepackage{amsmath}

\begin{document}

\title{Superradiance of a charged scalar field coupled to the Einstein-Maxwell
  equations}

\author{Olaf Baake}
\email{olaf.baake@aei.mpg.de}
\affiliation{Max Planck Institute for Gravitational Physics 
  (Albert Einstein Institute), Am M\"uhlenberg 1, 14476 Potsdam, Germany}
\author{Oliver Rinne}
\email{oliver.rinne@aei.mpg.de}
\affiliation{Hochschule f\"ur Technik und Wirtschaft Berlin,
  Treskowallee 8, 10318 Berlin, Germany}
\affiliation{Max Planck Institute for Gravitational Physics 
  (Albert Einstein Institute), Am M\"uhlenberg 1, 14476 Potsdam, Germany}

\date{October 26, 2016}

\begin{abstract}
  We consider the Einstein-Maxwell-Klein-Gordon equations for a spherically 
  symmetric scalar field scattering off a Reissner-Nordstr\"om black hole in 
  asymptotically flat spacetime.
  The equations are solved numerically using a hyperboloidal evolution scheme.
  For suitable frequencies of the initial data, superradiance is observed,
  leading to a substantial decrease of mass and charge of the black hole.
  We also derive a Bondi mass loss formula using the Kodama vector field and
  investigate the late-time decay of the scalar field.
\end{abstract}

\maketitle


\section{Introduction}
\label{s:intro}

Even though particles can never cross the event horizon from the inside of a 
black hole (at least classically), it is nevertheless possible to extract 
energy from a rotating or charged black hole.
For particles this can occur via the Penrose process \cite{Penrose1969}.
The field-theoretic analogue is superradiance, which involves the
scattering of scalar, electromagnetic or gravitational waves off a black hole. 
For a comprehensive recent review article of this field of research see
\cite{Brito2015}.

In general relativity, superradiance is mostly studied for rotating black holes
(the Kerr family of solutions) and is based on linear perturbation 
theory \cite{Press1973}.
It should be stressed that this involves a mode analysis.
There has been some controversy as to whether the predicted amplification 
factors are still meaningful for realistic wave packets.
Csizmadia \etal \cite{Csizmadia2013} reported an almost perfect reflection for 
a wave packet whose initial frequency content lies entirely in the superradiant 
regime.
Until recently little was known about superradiance in the nonlinear case.
East \etal \cite{East2014} presented fully nonlinear numerical evolutions of the vacuum Einstein equations for 
gravitational waves scattering off a rotating black hole and confirmed the
existence of superradiance.

Due to the lack of symmetries such simulations are very demanding.
A simpler model is obtained by considering a charged scalar field scattering
off a charged (Reissner-Nordstr\"om) black hole in spherical symmetry.
This 1+1-dimensional problem can be tackled with modest computational resources.
The test-field case (fixed black hole background spacetime) was studied by 
Di Menza and Nicholas \cite{DiMenza2015}.
The authors computed the frequency-dependent energy gain by constructing a 
conserved flux.
Such an approach is not feasible when the matter fields are coupled to the Einstein equations and 
the mass and charge of the black hole change during the superradiant scattering.
Instead it is the changes in those quantities themselves that need to be 
monitored, as in \cite{East2014} and in the present paper.
Numerical evolutions of the coupled Einstein-Maxwell-Klein-Gordon system were 
first carried out by Hod and Piran \cite{Hod1997} and more recently e.g.
by Torres and Alcubierre \cite{Torres2014},
however with a focus on gravitational collapse rather than superradiance.
Recently there has been an increasing interest in superradiance in the context
of the anti-de Sitter (AdS)--conformal field theory (CFT) correspondence.
An example is the numerical study of a superradiant instability of 
Reissner-Nordstr\"om-AdS black holes in \cite{Bosch2016}.
Due to the timelike infinity one effectively has reflective boundary conditions
that lead to an unbounded amplification of the superradiant waves.
A similar situation, namely a Reissner-Nordstr\"om black hole enclosed in a 
cavity, was investigated in \cite{Sanchis-Gual2016}.
To our best knowledge the present paper is the first study of superradiance 
in the coupled Einstein-Maxwell-Klein-Gordon system in asymptotically flat 
spacetimes.

Most numerical studies of asymptotically flat spacetimes truncate the 
computational domain at a finite distance from the black hole, where 
boundary conditions must be imposed. 
(This approach was taken e.g.~in \cite{DiMenza2015}.)
Spurious reflections of outgoing waves must be avoided on the relevant 
time scales. 
Designing such boundary conditions in the nonlinear case is highly non-trivial.
A more elegant solution is to incorporate future null infinity $\scri$ in the 
computational domain, e.g.~by foliating spacetime into hyperboloidal surfaces
that may be compactified towards future null infinity.
This also enables us to evaluate the Bondi mass and total charge at $\scri$
in a straightforward way.
We use a conformal 3+1 decomposition of the Einstein equations on hypersurfaces
of constant mean curvature (CMC) developed in \cite{Moncrief2009}.
Such CMC surfaces have the additional advantage that they are able to penetrate
black hole horizons, so that an excision boundary may be placed just inside
the horizon, where all characteristics leave the computational domain.
In \cite{Rinne2010} this approach was first implemented for the vacuum 
axisymmetric Einstein equations.
In \cite{Rinne2013} we included matter sources and studied late-time power-law
tails of matter fields in spherical symmetry.
The present formulation is based on \cite{Rinne2013}.

This paper is organized as follows.
In Sec.~\ref{s:formulation} we describe our formulation of the
Einstein-Maxwell-Klein-Gordon equations and the hyperboloidal evolution scheme.
We also investigate mass conservation with the help of the Kodama vector field
and derive a Bondi mass loss formula.
Details on redundant evolution equations and their regularity at $\scri$
and on alternative electromagnetic gauge conditions are deferred to  
Appendices \ref{s:einsteinevoln} and \ref{s:gaugefixing}.
Section \ref{s:evolution} is concerned with the numerical evolution of the 
system.
We describe our numerical methods, construction of initial data, and various
notions of charge and mass used to analyze the results.
After performing a number of code tests, we present our main results on the
occurrence and amount of superradiance depending on the chosen parameters.
A separate subsection is devoted to the analysis of the late-time decay of the
scalar field.
We conclude in Sec.~\ref{s:concl}.


\section{Formulation and theoretical analysis}
\label{s:formulation}

In this section we describe our formulation of the Einstein-Maxwell-Klein-Gordon
equations and their reduction to spherical symmetry. 
Our gauge choices for the Einstein and Maxwell equations are explained,
in particular our use of constant-mean-curvature slices extending to future
null infinity.
In Sec.~\ref{s:masscons} we investigate the Kodama vector field, construct a 
conserved flux and relate it to the Hawking mass.
This enables us to derive a Bondi mass loss formula.


\subsection{Field equations}
\label{s:fieldeqns}

We consider a massive charged complex scalar field minimally coupled to the 
Einstein-Maxwell equations.
The action is given by
\begin{equation}
  \label{e:action}
  \begin{split}
    S =& \int \diff^4x \, \mu_{\four g} \left( \frac{1}{16\pi} \left[ \four R 
        - F_{\mu \nu} F^{\mu \nu}  \right] \right. \\
    & \left. - \frac{1}{2} \left[ \four g^{\mu\nu} \left(\gaugeD_{\mu} \phi \right)^* \left(\gaugeD_{\nu} \phi \right) + m^2 |\phi|^2 \right] \right),
  \end{split}
\end{equation}
where $\four g_{\mu\nu}$ is the spacetime metric, $\mu_{\four g}$ its volume 
element and $\four R$ the scalar curvature.
We use geometric and Gaussian units.
In terms of the vector potential $A_\mu$, the Maxwell field strength tensor is
\begin{equation}
  \label{e:Fdef}
  F_{\mu\nu} = \partial_{\mu} A_{\nu} - \partial_{\nu} A_{\mu}.
\end{equation}
The gauge-covariant derivative is defined as
\begin{equation}
  \gaugeD_{\mu} := \four \nabla_{\mu} + iq A_{\mu},
\end{equation}
where $\four \nabla$ is the covariant derivative of $\four g_{\mu\nu}$.
The mass and charge of the scalar field are $m$ and $q$ respectively.

We perform a conformal transformation
\begin{equation}
  \label{e:fourg1}
  \four g_{\mu\nu} = \Omega^{-2} \four \gamma_{\mu\nu},
\end{equation}
where the conformal factor $\Omega \searrow 0$ at $\scri$.
In spherical symmetry we may write the conformal metric in isotropic 
coordinates as 
\begin{equation}
  \label{e:fourg2}
  \four \gamma = -\tN^2 \rmd t^2 + (\rmd r + rX \, \rmd t)^2 + r^2 d \sigma^2
\end{equation}
with $d\sigma^2 = \rmd\theta^2 + \sin^2\theta \, \rmd\phi^2$.
We consider an ADM decomposition \cite{Arnowitt1962} with respect to the time 
coordinate $t$. 
Constant mean curvature (CMC) slicing is used; i.e., the mean curvature of the
$t=\const$ slices is a spacetime constant $K>0$.
The tracefree part of the ADM momentum $\pi^{\tr ij}$ (cf.~\cite{Moncrief2009}) 
has only one degree of freedom in spherical symmetry, which we take to be
$\Pi := (r^4 \sin\theta)^{-1} \pi^{\tr rr}$.
The gravitational field is thus described by the four variables 
$\Omega, \tN, X$ and $\Pi$, which are functions of $t$ and $r$ only.
In the following we use an overdot to denote $t$-derivatives and
a prime to denote $r$-derivatives.

Preserving the isotropic spatial coordinate condition and the CMC slicing 
condition under the time evolution yields
\begin{eqnarray}
  \label{e:isotropic}
  0 &=& r^{-1} X' + \tfrac{3}{2} \tN \Pi,\\
  \label{e:cmc}
  0&=&-\Omega^2 \tN'' + 3 \Omega\Omega'\tN' - 2\Omega^2 r^{-1}\tN' 
  - \tfrac{3}{2}\Omega'^2 \tN   \nonumber\\
  &&+ \sixth \tN K^2 + \tfrac{15}{8} \tN \Omega^2 r^4 \Pi^2
   + 4\pi \tN \Omega^4 (\tilde S + 2 \tilde \rho).
\end{eqnarray}
In spherical symmetry, the geometry is completely determined by the
Einstein constraint equations,
\begin{eqnarray}
  \label{e:hamcons}
  0 &=& -4 \Omega \Omega'' + 6 \Omega'^2 
  - 8 \Omega r^{-1}\Omega' + \tfrac{3}{2} \Omega^2 r^4 \Pi^2 \nonumber\\
  &&- \tfrac{2}{3} K^2 + 16\pi \Omega^4 \tilde \rho, \\
  \label{e:momcons}
  0 &=& \Omega (r\Pi' + 5 \Pi) - 2r\Omega' \Pi 
  + 8\pi \Omega^3 r^{-1} \tilde J^r.
\end{eqnarray}
The source terms $\tilde \rho$, $\tilde S$ and $\tilde J^r$ in 
\eqref{e:cmc}--\eqref{e:momcons} are components
of the conformally rescaled energy-momentum tensor
$\tilde T_{\mu\nu} = \Omega^{-2} T_{\mu\nu}$ and are defined for our matter model
below in \eqref{e:31rho}--\eqref{e:31S}.
There are redundant evolution equations for $\Omega$ and $\Pi$, given in
Appendix \ref{s:einsteinevoln}, that we monitor
during the evolution in order to check the accuracy of our code.

The Maxwell equations are conformally invariant and hence we may define the
fields in terms of the conformal metric $\four \gamma_{\mu\nu}$,
indicated by tildes in the following.
In spherical symmetry the vector potential may be written as
\begin{equation}
  \tilde A = - \tN \tilde\Phi \, \rmd t + \tilde a_r (\rmd r + r X \,\rmd t) ,
\end{equation}
and the field strength tensor
\begin{equation}
  \label{e:Fansatz}
  \tilde F = - \tN \tilde E_r \, \rmd t \wedge \rmd r.
\end{equation}
We may impose one gauge condition on the vector potential, and the one we 
choose is temporal gauge 
\begin{equation}
  \label{e:gauge}
  \tilde{\Phi} = 0.
\end{equation}
Alternative gauge conditions are discussed in Appendix \ref{s:gaugefixing}.

The definition \eqref{e:Fdef} of the field strength tensor in terms of the 
vector potential implies an evolution equation for $\tilde a^r$:
\begin{equation}
  \label{e:eoma} 
  \dot{\tilde{a}}^r = \left(r X \tilde{a}^r\right)' - \tilde{N}\tilde{E}^r .
\end{equation}
The Maxwell equations imply an evolution equation for the electric field
\begin{equation}
  \label{e:eomE} 
  \dot{\tilde{E}}\,^r = r X \tilde{E}^r{}' + 2 X \tilde{E}^r - r \tilde{E}^r X' 
  - \frac{3}{2} r^2 \tilde{N}\tilde{E}^r \Pi 
  - 4\pi\tilde{N}\tilde{j}^r_{\mathrm{e}}
\end{equation}
and the Gauss constraint
\begin{equation}
  \label{e:gaucons}
  \left( r^2 \tilde{E}^r \right)' = 4\pi r^2 \tilde{\rho}_{\mathrm{e}}.
\end{equation}
The source terms $\tilde{\rho}_{\mathrm{e}}$ and 
$\tilde{j}^r_{\mathrm{e}}$ are given below in 
\eqref{e:31rhoe}--\eqref{e:31je}.
We normally use \eqref{e:gaucons} to solve for $\tilde{E}^r$ and monitor 
\eqref{e:eomE} during the evolution but we have checked that the 
opposite choice gives identical results within numerical accuracy.

We introduce a conformally rescaled scalar field
\begin{equation}
  \tilde \phi := \Omega^{-1} \phi
\end{equation}
and write the Klein-Gordon equation in first-order form 
(in time) by introducing a new variable $\tilde \psi$:
\begin{eqnarray}
  \label{e:eomphi}
  \dot{\tilde{\phi}} &=& r X \tilde{\phi}' + X \tilde{\phi} 
  + \tilde{N} \left( \tilde{\psi} - \frac{1}{2}r^2 \tilde{\phi}\Pi \right), \\
  \label{e:eompsi}
  \dot{\tilde{\psi}} &=& r X \tilde{\psi}'+ \tilde{N} \left( \tilde{\phi}'' 
  + 2r^{-1}\tilde{\phi}' - \frac{1}{4} \tilde{\phi} r^4 \Pi^2 \right) 
  + \tilde{\phi}' \tilde{N}' \nonumber\\
  && + \tilde{\psi} \left( -\tilde{N}r^2 \Pi + 2 X \right) 
  - \Omega^{-2} m^2 \tilde{N}\tilde{\phi} + iq r^{-2} \tilde{\phi} 
  \left( r^2 \tN \tilde{a}^r \right)' \nonumber\\
  && + \frac{1}{3} \tilde{\phi} \left( \tilde{N}'' + 2 r^{-1}\tilde{N}' \right) 
  - q^2 \tilde{\phi} \tilde{N} \left( (\tilde{a}^r)^2 - \tilde{\Phi}^2 \right) 
  \nonumber\\
  && + 2iq \tilde{N} \left[ \tilde{a}^r \tilde{\phi}' 
  + \tilde{\Phi} \tilde{\psi} - \frac{1}{2} r^2 \Pi \tilde{\Phi} \tilde{\phi} 
  + \tilde{N}^{-1} X \tilde{\Phi} \tilde{\phi} \right] \nonumber\\
  && - \frac{4}{3}\pi \Omega^{2} \tilde{N}\tilde{\phi} \left[ 
  iq\tilde{a}^r\left( \tilde{\phi}^* \tilde{\phi}' 
  - \tilde{\phi} \tilde{\phi}'^* \right) - |\tilde{\phi}'|^2 
  - q^2|\tilde{\phi}|^2 (\tilde{a}^r)^2 \right. \nonumber\\
  && \left. \qquad\qquad - iq \tilde{\Phi} \left( \tilde{\phi} \tilde{\psi}^* 
  - \tilde{\phi}^* \tilde{\psi} \right) + |\tilde{\psi}|^2 
  + q^2|\tilde{\phi}|^2 \tilde{\Phi}^2 \right] \nonumber\\
  && + \frac{4}{3}\pi \Omega \tilde{N}\tilde{\phi} \left[ \frac{1}{3} K 
  \left( \tilde{\psi}\tilde{\phi}^* + \tilde{\psi}^* \tilde{\phi} \right) 
  + \Omega' \left( \tilde{\phi}^* \tilde{\phi}' + \tilde{\phi} \tilde{\phi}'^* 
  \right) \right] \nonumber\\
  && - \frac{4}{3}\pi \tilde{N}\tilde{\phi} \left[ \frac{1}{9} 
  |\tilde{\phi}|^2 K^2 - |\tilde{\phi}|^2 \Omega'^2 - 2 m^2 |\tilde{\phi}|^2 
  \right].
\end{eqnarray}
The somewhat non-standard definition of $\tilde\psi$ given by \eqref{e:eomphi}
is used in order to avoid time derivatives of the lapse and shift in its 
evolution equation \eqref{e:eompsi}, cf.~\cite{Rinne2013}.
In deriving \eqref{e:eompsi}, we have re-expressed the four-dimensional
scalar curvature in terms of the trace of the energy-momentum tensor
using the Einstein equations, which produces terms quadratic in the scalar 
field.
This is required in order to remove a term containing a negative power of the 
conformal factor, which would be formally singular at $\scri$.
In \cite{Rinne2013} a different approach based on conformal rather 
than minimal coupling of the scalar field is taken.

Numerically, we have found it to be advantageous to also perform a first-order
reduction in space by introducing a new variable 
$\tilde\xi := r^{-1} \tilde\phi'$, 
and we evolve the real and imaginary part of the complex scalar field variables 
separately.

The source terms appearing in the slicing condition \eqref{e:cmc} and the
Einstein constraint equations \eqref{e:hamcons}--\eqref{e:momcons} are
\begin{eqnarray}
  \label{e:31rho}
  \tilde{\rho} &=& \frac{1}{2} \left( |\tilde{\phi}'|^2 
  + |\tilde{\psi}|^2 \right) + \frac{1}{18} \Omega^{-2} |\tilde{\phi}|^2 K^2 
  \nonumber\\
  && + \frac{1}{2} \left[ iq \tilde{a}^r \left( \tilde{\phi}\tilde{\phi}'^* 
  - \tilde{\phi}^*\tilde{\phi}' \right) + q^2 |\tilde{\phi}|^2 \left( 
  (\tilde{a}^r)^2 + \tilde{\Phi}^2 \right) \right] \nonumber\\
  && + \frac{1}{2} \Omega^{-1} \Omega' \left( \tilde{\phi}^* \tilde{\phi}' 
  + \tilde{\phi} \tilde{\phi}'^* \right) + \frac{1}{2} \Omega^{-2} 
  |\tilde{\phi}|^2 \Omega'^2 \nonumber\\
  && + \frac{1}{2} iq \tilde{\Phi} \left( \tilde{\phi}^* \tilde{\psi} 
  - \tilde{\phi} \tilde{\psi}^* \right) - \frac{1}{6} \Omega^{-1} K \left( 
  \tilde{\phi}^* \tilde{\psi} + \tilde{\phi} \tilde{\psi}^* \right) \nonumber\\
  && + \frac{1}{8\pi} ( \tilde{E}^r)^2 + \frac{1}{2}\Omega^{-2} m^2 
  |\tilde{\phi}|^2,\\
  \label{e:31J}
  \tilde{J}^r &=& q^2 |\tilde{\phi}|^2 \tilde{\Phi} \tilde{a}^r 
  + \frac{1}{2}iq \tilde{\Phi} \left( \tilde{\phi}\tilde{\phi}'^* 
  - \tilde{\phi}^*\tilde{\phi}' \right) + \frac{1}{3} \Omega^{-2} 
  |\tilde{\phi}|^2 \Omega' K \nonumber\\
  && - \frac{1}{2} \left[ \tilde{\phi}'^* \tilde{\psi} 
  + \tilde{\phi}' \tilde{\psi}^* + iq \tilde{a}^r \left( \tilde{\phi}
  \tilde{\psi}^* - \tilde{\phi}^*\tilde{\psi} \right) \right] \nonumber\\
  && + \frac{1}{2} \Omega^{-1} \left[ \frac{1}{3} K \left( \tilde{\phi}^* 
  \tilde{\phi}' + \tilde{\phi} \tilde{\phi}'^* \right) - \Omega' \left( 
  \tilde{\phi}^* \tilde{\psi} + \tilde{\phi} \tilde{\psi}^* \right) \right],
  \nonumber\\ \\
  \label{e:31S}
  \tilde{S} &=& -\frac{1}{2}|\tilde{\phi}'|^2-\frac{iq}{2}\tilde{a}^r \left( 
  \tilde{\phi}\tilde{\phi}'^* - \tilde{\phi}^*\tilde{\phi}' \right) 
  - \frac{q^2}{2}|\tilde{\phi}|^2 (\tilde{a}^r)^2 \nonumber\\
  && - \frac{1}{2}\Omega^{-1}\Omega' \left( \tilde{\phi}^*\tilde{\phi}' 
  + \tilde{\phi}\tilde{\phi}'^* \right) - \frac{1}{2}\Omega^{-2} 
  |\tilde{\phi}|^2 \Omega'^2 \nonumber\\
  && + \frac{3}{2}\left[ |\tilde{\psi}|^2 + iq \tilde{\Phi}\left( 
  \tilde{\phi}^* \tilde{\psi} - \tilde{\phi}\tilde{\psi}^* \right) 
   + q^2 |\tilde{\phi}|^2 \tilde{\Phi}^2 \right] \nonumber\\
  && - \frac{1}{2}\Omega^{-1}K \left( \tilde{\phi}^*\tilde{\psi} 
  + \tilde{\phi}\tilde{\psi}^* \right) + \frac{1}{6}\Omega^{-2} 
  |\tilde{\phi}|^2 K^2 \nonumber\\
  && + \frac{1}{8\pi} ( \tilde{E}^r )^2 - \frac{3}{2}\Omega^{-2} m^2 
  |\tilde{\phi}|^2.
\end{eqnarray}
The source terms appearing in the Maxwell equations 
\eqref{e:eomE}--\eqref{e:gaucons} are
\begin{eqnarray}
  \label{e:31rhoe}
  \tilde{\rho}_{\mathrm{e}} &=& -\frac{iq}{2}\left( \tilde{\phi}^* 
  \tilde{\psi} - \tilde{\phi}\tilde{\psi}^* \right) - q^2 |\tilde{\phi}|^2 
  \tilde{\Phi},\\
  \label{e:31je}
  \tilde{j}^r_{\mathrm{e}} &=& \frac{iq}{2} \left( \tilde{\phi}^* 
  \tilde{\phi}' - \tilde{\phi} \tilde{\phi}'^* + 2iq |\tilde{\phi}|^2 
  \tilde{a}^r \right).
\end{eqnarray}
%


\subsection{Mass conservation}
\label{s:masscons}

In the dynamical spacetimes we study, there is no timelike Killing vector field.
Nevertheless, in spherical symmetry one can construct a preferred timelike
vector field \cite{Kodama1980,Racz2006,Csizmadia2010}.
For a general spherically symmetric metric of the form
\begin{equation}
  ds^2 = \two g_{ab} \, \rmd x^a \rmd x^b 
  + \bar r^2 (\rmd\theta^2 + \sin^2\theta \, \rmd\phi^2),
\end{equation}
with a two-dimensional Lorentzian metric $\two g_{ab}$ (indices $a,b$ ranging 
over $t,r$), this \emph{Kodama vector field} is given by
\begin{equation}
  K^\mu = \two \epsilon^{\mu\nu} \partial_\nu \bar r,
\end{equation}
where $\two \epsilon$ is the volume element of $\two g$.
For our form of the metric \eqref{e:fourg1}--\eqref{e:fourg2}, we obtain
\begin{eqnarray}
  K^t &=& \tN^{-1} (r \Omega' - \Omega), \\
  K^r &=& -r \tN^{-1} \dot \Omega, \\
  K^\theta &=& K^\phi = 0.
\end{eqnarray}

The Kodama vector field has the remarkable properties
\begin{equation}
  \four \nabla^\mu K_\mu = 0
\end{equation}
and
\begin{equation}
  \four G^{\mu\nu} \four\nabla_\mu K_\nu = 0.
\end{equation}
Together with Einstein's equations, the latter implies that the current
\begin{equation}
  J_\mathrm{K}^\mu := T^{\mu\nu} K_\nu 
\end{equation}
is conserved, $\four \nabla_\mu J_\mathrm{K}^\mu = 0$.
In our formulation we obtain
\begin{eqnarray}
  J_\mathrm{K}^t &=& \Omega^4 \left[ \Omega^2 \tN^{-1} \tilde J^r K_r 
    - \tilde \rho K^t \right], \\
  J_\mathrm{K}^r &=& \Omega^4 \left[ \Omega^2 \left( 
      \tilde S^{\textrm{\scriptsize tr}\,rr}
     + \third \tilde S  - r X \tN^{-1} \tilde J^r \right) K_r \right.\nonumber\\
      && \qquad \left. + \left( r X \tilde \rho - \tN \tilde J^r \right) K^t 
      \right], \\
  J_\mathrm{K}^\theta &=& J_\mathrm{K}^\phi = 0.
\end{eqnarray}
(The source term $\tilde S^{\textrm{\scriptsize tr}\,rr}$ is defined in
\eqref{e:31Strrr}.)

The Kodama vector field is closely related to the Misner-Sharp or Hawking mass \cite{Csizmadia2010}
\begin{equation}
  \label{e:Hawkm}
  M_{\mathrm{H}} = \frac{1}{2}\frac{r}{\Omega} \left[ 1 + g_{\mu\nu}K^\mu K^\nu
  \right].
\end{equation}
Its derivatives turn out to be
\begin{equation}
  \nabla_\mu M_{\mathrm{H}} = \frac{8\pi r^2 \tilde{N}}{2 \Omega^4} 
  \left( -J^r_{\mathrm{K}}, J^t_{\mathrm{K}}, 0, 0 \right), 
\end{equation}
so that the integral of the Kodama flux is essentially given by the difference
of the Hawking masses at the ends:
\begin{eqnarray}
  \label{e:tflux}
  4\pi \int_{r_0}^{r_1} r^2 \tN \Omega^{-4} J_\mathrm{K}^t \, \rmd r
  &=& M_{\mathrm{H}}(t,r_1) - M_{\mathrm{H}}(t,r_0),  \\
  \label{e:rflux}
  4\pi r^2 \int_{t_0}^{t_1} \tN \Omega^{-4} J_\mathrm{K}^r \, \rmd t 
  &=& M_{\mathrm{H}}(t_0,r) - M_{\mathrm{H}}(t_1,r).
\end{eqnarray}
Taking the limit of \eqref{e:rflux} at $\scri$, where the Hawking mass coincides
with the Bondi mass $M_\mathrm{B}$, and using the regularity conditions stated
in Appendix \ref{s:einsteinevoln}, a lengthy calculation results in the Bondi 
mass loss formula (see also \cite{Scholtz2014})
\begin{equation}
  \label{e:Bmloss}
   \partial_t M_{\mathrm{B}} = -4\pi r^2 \left| \left\{ \partial_t 
  - \frac{i}{3}rqK \left( \tilde{\Phi} + \tilde{a}^r \right) \right\}
  \tilde{\phi} \right|_{\scri}^2,
\end{equation}
which is manifestly non-positive.


\section{Numerical evolution}
\label{s:evolution}

In this section we present our numerical evolutions of the system derived 
in the previous section.
In Sec.~\ref{s:nummethod} we briefly describe the numerical methods we use.
The initial data for a Reissner-Nordstr\"om black hole with scalar field
perturbation are constructed in Sec.~\ref{s:ini}.
In Sec.~\ref{s:chargemass}, various notions of charge and mass are introduced,
which are needed to evaluate our numerical evolutions.
In Sec.~\ref{s:tests} we perform mass/charge conservation and convergence tests
in order to check the accuracy of our code.
Sec.~\ref{s:numresults} contains our main results: evolutions for various
choices of parameters are presented and evaluated with regard to the 
existence and the amount of superradiance. 
Finally in Sec.~\ref{s:qnmtails}, we investigate the late-time behavior of the
scalar field and compare with known perturbative results on quasi-normal modes
and power-law tails.


\subsection{Numerical method}
\label{s:nummethod}

We discretize the equations in space using fourth-order finite differences.
A mapping of the radial coordinate with an adjustable 
parameter \cite{Rinne2013} is used
in order to provide more resolution where it is needed, especially near the
black hole horizon where the fields typically develop steep gradients.
The outermost grid point is placed at $\scri$, which we choose to correspond
to $r=1$.
Typical resolutions used for the simulations in this paper range 
from $2000$ to $10000$ radial grid points.

Following the method of lines, the evolution equations  \eqref{e:eoma},
\eqref{e:eomphi} and \eqref{e:eompsi} are first discretized in space and then 
integrated forward in time using a fourth-order Runge-Kutta method with 
sixth-order Kreiss-Oliger dissipation \cite{Kreiss1973}.
At each time step, the ODEs \eqref{e:isotropic}--\eqref{e:momcons} 
and \eqref{e:gaucons} are solved using a Newton-Raphson method, at each 
iteration solving the resulting linear system using a direct band-diagonal 
solver.
(Alternatively, as mentioned in Sec.~\ref{s:fieldeqns}, we may replace 
the Gauss constraint \eqref{e:gaucons} with the evolution equation 
\eqref{e:eomE} for the electric field, which yields identical results
within numerical accuracy.)

Our treatment of the boundaries follows \cite{Rinne2013}.
We place an inner excision boundary just inside the apparent horizon of the 
black hole, whose location is determined at each time step as the outermost zero
of the expansion of outgoing null rays,
\begin{equation}
  \label{e:expansion}
  \theta_+ (r) = \third K - \half \Omega r^2 \Pi + r^{-1} \Omega - \Omega'.
\end{equation}
One-sided finite differences are used at the excision boundary.
Since this boundary lies inside the black hole, all characteristics 
leave the domain and hence no boundary conditions are required for the 
evolution equations.
We choose to freeze the conformal lapse $\tilde N$ at the inner boundary,
yielding a Dirichlet boundary condition for the slicing condition \eqref{e:cmc}.
Inner Dirichlet boundary conditions for the Einstein constraint equations 
\eqref{e:hamcons}--\eqref{e:momcons} and the Gauss constraint \eqref{e:gaucons}
are obtained by evolving $\Omega$, $\Pi$ and $\tilde E^r$ there according to 
their evolution equations \eqref{e:dtOmega}--\eqref{e:dtpi} and \eqref{e:eomE}.

The outer boundary $\scri$ is an outflow boundary and hence we use one-sided 
differences there as well, with no boundary conditions for the
evolution equations.
The conformal lapse is set to $\tilde N = \third K r$ at $\scri$, which
ensures that our time coordinate $t$ coincides with Bondi time \cite{Rinne2013}.
Outer Dirichlet boundary conditions on $X$, $\Omega$ and $\Pi$ 
follow from the regularity conditions at $\scri$, equations
\eqref{e:scrO}, \eqref{e:scrpi} and \eqref{e:scrX}.

In all our evolutions the value of the mean curvature is taken to be $K=1/2$.

The code has been written in and the figures produced with 
\texttt{Python}, making use of the \texttt{NumPy}, \texttt{SciPy} and
\texttt{matplotlib} extensions.


\subsection{Initial data}
\label{s:ini}

We choose initial data that is close to the Reissner-Nordstr\"om spacetime. 
First the geometry variables ($\Omega$, $\Pi$, $\tilde{N}$ and $X$) are 
set to coincide with this solution, then the initial data for the scalar field 
are specified and finally the constraints and elliptic gauge conditions are 
re-solved for the geometry variables.

The Reissner-Nordstr\"om spacetime in (uncompactified) CMC coordinates is 
given by \cite{Brill1980}
\begin{eqnarray}
  \label{e:cmcreinor}
    \four g &=& - \left(1 - \frac{2M}{\bar{r}} 
    + \frac{Q^{2}}{\bar{r}^2}\right) \rmd t^2 
    + \frac{1}{f^2} \rmd\bar{r}^2 \\
    && - \frac{2 a}{f} \rmd t \, \rmd\bar{r} 
    + \bar{r}^2 \left( \rmd\theta^2 
    + \sin^2\theta\,\rmd\varphi^2 \right),
\end{eqnarray}
with
\begin{eqnarray}
  \label{e:cmcfunctions}
  f(\bar{r}) &=& \left( 1 - \frac{2M}{\bar{r}} + \frac{Q^{2}}{\bar{r}^2} 
  + a^2 \right)^{1/2},\\
  \label{e:cmcfunctions2}
  a(\bar{r}) &=& \frac{K \bar{r}}{3} - \frac{H}{\bar{r}^2},
\end{eqnarray}
where $M$ (mass), $K$ (mean curvature) and $H$ are constants. 
We transform the radial coordinate $\bar{r}$ to a new radial coordinate $r$
by demanding that the spatial metric be manifestly conformally flat. 
For convenience we work with $s := 1/\bar{r}$ due to its finite range, 
which yields the ODE
\begin{equation}
  \label{e:isoode}
  \frac{\rmd s}{\rmd r} = \frac{-F(s)^{1/2}}{r},
\end{equation}
with 
\begin{eqnarray}
  F(s) &=& s^2 - 2 M s^3 + Q^2 s^4 + A(s)^2, \\
  A(s) &=& \frac{K}{3} - H s^3,
\end{eqnarray}
and we choose $r=1$ to correspond to $\scri$ ($s = 0$). 
We obtain the $r$-coordinate of the horizon by numerically integrating
\begin{equation}
  \label{e:rhorizon}
  r_{\mathrm{h}} = \exp \left( - \int^{s_{\mathrm{h}}}_0 F(s)^{-1/2} \,
  \rmd s \right),
\end{equation}
where $s_{\mathrm{h}} = 1/\bar r_+$, and
\begin{equation}
  \bar r_\pm = M \pm \sqrt{M^2 - Q^2}
\end{equation}
are radii of the outer and inner horizon in Schwarzschild coordinates.
We place the excision boundary just inside the outer horizon: 
$r_{\mathrm{min}} = \alpha \, r_{\mathrm{h}}$, where 
typical values of $\alpha$ are between 0.8 and 0.9.
The ODE \eqref{e:isoode} is then solved numerically on the interval 
$r \in \left[ r_{\mathrm{min}} , 1 \right]$ with the initial condition 
$s(1) = 0$.

The geometry variables can be expressed in terms of the numerically determined 
function $s(r)$ as
\begin{eqnarray}
  \Omega &=& rs, \label{e:iniO} \\
  \Pi &=& 2H s^2 r^{-3}, \label{e:inipi} \\
  \tilde{N} &=& r F(s)^{1/2}, \label{e:iniN} \\
  X &=& H s^3 - \third K. \label{e:iniX}
\end{eqnarray}

For the scalar field we choose initial data that are supported
sufficiently far outside the black hole, where the background is almost flat. 
We take an ingoing solution of the wave equation on a Minkowski background:
\begin{equation}
  \label{e:iniphi}
  \tphi (\bar{t}, \bar{r}) = \frac{A}{\bar{r}} \exp \left( i \omega
    (\bar{r} + \bar{t}) - \frac{(\bar{r} - \bar{r}_0 +
      \bar{t})^2}{\bar \sigma^2} \right).
\end{equation}
This is then expressed in terms of our CMC coordinates $t, r$, which are related
to the standard Minkowski coordinates $\bar t, \bar r$ via
\begin{eqnarray}
  \label{e:minkt}\bar t &=& t + \left( \frac{3}{K} \right) 
  \left( \frac{1 + r^2}{1 - r^2} \right), \\
  \label{e:minkr}
  \bar r &=& \frac{6 r}{K\left( 1 - r^2 \right)}.
\end{eqnarray}
Now we choose $\bar r_0$ such that the wave packet is localized sufficiently 
far outside the black hole 
(we use $\bar r_0=20M$, which corresponds to $r_0= 0.55$) 
and take $\tphi(t=0,r)$ as initial data for the scalar field.

The initial data for the electromagnetic field have to correspond to the 
Reissner-Nordstr\"om spacetime. 
Together with the gauge condition \eqref{e:gauge} we set
\begin{equation}
  \label{e:iniem}
  \tilde{E}^r = \frac{Q}{r^2}, \quad  \tilde{a}^r = \frac{Q}{ar}.
\end{equation}
Alternative gauge conditions (and our reasons for not using them)
are discussed in Appendix \ref{s:gaugefixing}.


\subsection{Charge and mass}
\label{s:chargemass}

As in \cite{Torres2014} we introduce the charge $Q(r)$ inside a sphere of 
radius $r$:
\begin{equation}
  \label{e:Qint}
  Q(r) = \int_{S(\bar{r})} \rho_e \rmd V = \int_{S(r)} \tilde{\rho}_e 
  \rmd\tilde{V},
\end{equation}
which using the Gauss constraint \eqref{e:gaucons} can be written as
\begin{equation}
  \label{e:Qr}
  Q(r) = r^2 \tilde{E}^r(r).
\end{equation}

Let us first consider a static black hole. 
Its irreducible mass is given by the area of the event horizon $A_{\mathrm{h}}$:
\begin{equation}
  \label{e:mirr}
  M_{\mathrm{irr}} = \sqrt{\frac{A_{\mathrm{h}}}{16 \pi}} = M_\mathrm{H} (r_\mathrm{h}).
\end{equation}
The mass of the black hole can now be calculated as
\begin{equation}
  \label{e:BHmass}
  M_{\mathrm{BH}} = M_{\mathrm{irr}} 
  + \frac{Q(r_{\mathrm{h}})^2}{4 M_{\mathrm{irr}}}.
\end{equation}
These formulas are valid if spacetime is static in a neighbourhood of the 
horizon. 
This is the case for the final state of the evolution as well as
for the initial data, which consist of a scalar field pulse
supported far away from the horizon.
During the evolution, we still compute $M_{\mathrm{irr}}$ and $M_{\mathrm{BH}}$
according to the above expressions, but evaluated at the \emph{apparent} 
horizon.
The second equality in \eqref{e:mirr} still holds if h refers to the apparent 
horizon, but this differs slightly from the event horizon during the dynamical 
phase of the evolution.
Hence when we plot \eqref{e:mirr} and \eqref{e:BHmass} as functions of time,
they cannot strictly be interpreted as the irreducible mass and black hole mass
during the dynamical phase but its initial and final values are correct.

Analogously to the rotational energy in \cite{East2014}, we define the charge 
energy of the black hole as
\begin{equation}
  \label{e:chargeE}
  E_{\mathrm{Q}} := M_{\mathrm{BH}} - M_{\mathrm{irr}}.
\end{equation}
During superradiant scattering, both the charge energy and the mass of the black
hole decrease:
\begin{equation}
  \Delta E_{\mathrm{Q}} = \underbrace{M_{\mathrm{BH}}^\mathrm{final} 
    - M_{\mathrm{BH}}^\mathrm{initial}}_{\Delta M_\mathrm{BH} \leqslant 0}
  - \underbrace{(M_\mathrm{irr}^\mathrm{final} 
    - M_\mathrm{irr}^\mathrm{initial} )}_{\Delta M_\mathrm{irr} \geqslant 0} 
  \leqslant 0.
\end{equation}
Part of the charge energy is carried away by the wave, part of it increases
the irreducible mass and hence is no longer extractable.
Following \cite{East2014} we define the efficiency of this process as
\begin{equation}
  \label{e:effi}
  \eta := \frac{\Delta M_{\mathrm{BH}}}{\Delta E_{\mathrm{Q}}}.
\end{equation}

Alternatively, we may compare the energy of the initial scalar field pulse
\begin{equation}
  E_\phi^\mathrm{initial} = 
  M_\mathrm{B}^\mathrm{initial} - M_\mathrm{BH}^\mathrm{initial}.
\end{equation}  
with the total energy radiated away at infinity,
\begin{equation}
  -\Delta M_\mathrm{B} = M_\mathrm{B}^\mathrm{initial} -
  M_\mathrm{B}^\mathrm{final} > 0.
\end{equation}  
From these quantities we define an alternative efficiency
\begin{equation}
  \label{e:effi2}
  \hat \eta := \frac{-\Delta M_\mathrm{B}}{E_\phi^\mathrm{initial}} - 1.
\end{equation}


\subsection{Code tests}
\label{s:tests}

The first test of our code consists in the evolution of the Reissner-Nordstr\"om
background spacetime, without any scalar field.
On the timescale relevant for the scalar field scattering experiments presented 
later ($t \lesssim 60 M$), mass (Bondi, black hole and irreducible)
is conserved to a relative error of $2\times10^{-6}$
and charge (at future null infinity and at the horizon) is conserved to a 
relative error of $10^{-5}$.
Both are negligible compared with the observed changes of charge and mass 
during scalar field scattering.

\begin{figure}
  \centerline{\includegraphics[width=.475\textwidth]{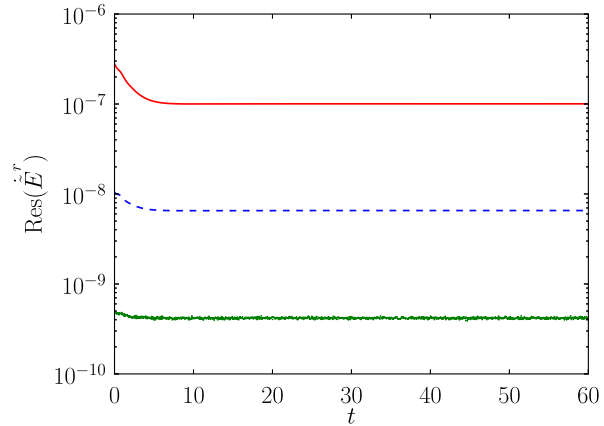}}
  \caption{\label{f:convtest}
    Convergence test for three different resolutions: $1000$ (solid red), 
    $2000$ (dashed blue) and $4000$ (noisy green) grid points.
    Shown is the residual of the evolution equation \eqref{e:eomE} of the 
    electric field as a function of time.
    The parameters used here are
    $M = 1$, $Q = 0.282$, $m = 0$, $q = 0.282$, $\omega = 0.5$, $\bar r_0 = 20$,
    $\bar \sigma = 0.5$ and $A = 0.01$.
  }
\end{figure}

Next we perform a convergence test for an evolution with scalar field.
Figure \ref{f:convtest} shows the residual of the evolution equation 
\eqref{e:eomE} for $\tilde E^r$, which we do not impose actively as
we solve the Gauss constraint \eqref{e:gaucons} for $\tilde E^r$.
(To compute the time derivative in the evolution equation numerically,
we use fourth-order finite differences, using data from five subsequent time
levels.)
The observed decrease of the residual for successively doubled resolutions
is close to the expected value of $2^4 = 16$ for a fourth-order accurate 
finite-difference method.
Similar plots are obtained for the residuals of the evolution equations 
\eqref{e:dtOmega} and \eqref{e:dtpi} for $\Omega$ and $\Pi$.


\subsection{Numerical results}
\label{s:numresults}

\begin{figure}
  \centerline{\includegraphics[width=.475\textwidth]{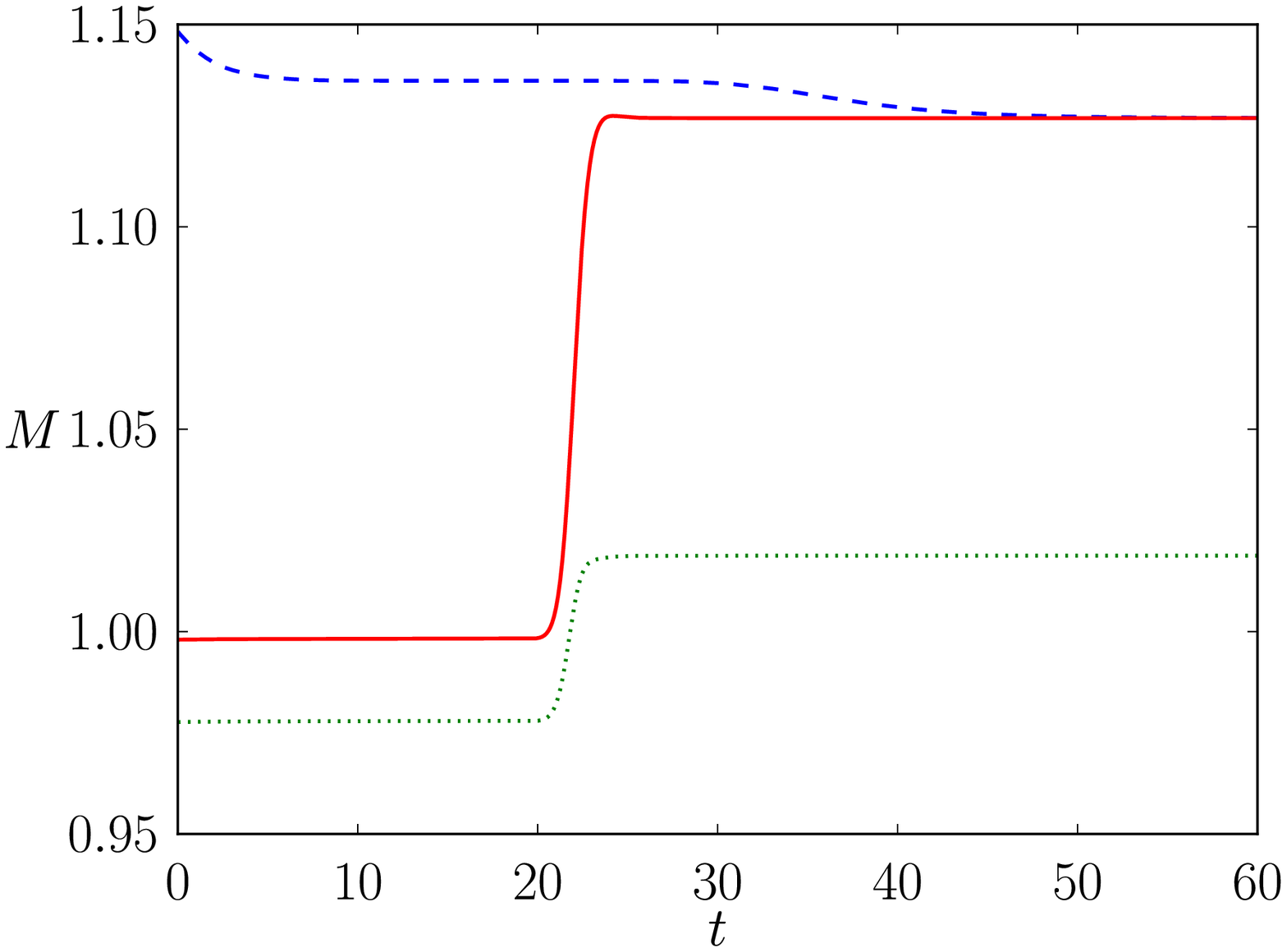}}
  \centerline{\includegraphics[width=.475\textwidth]{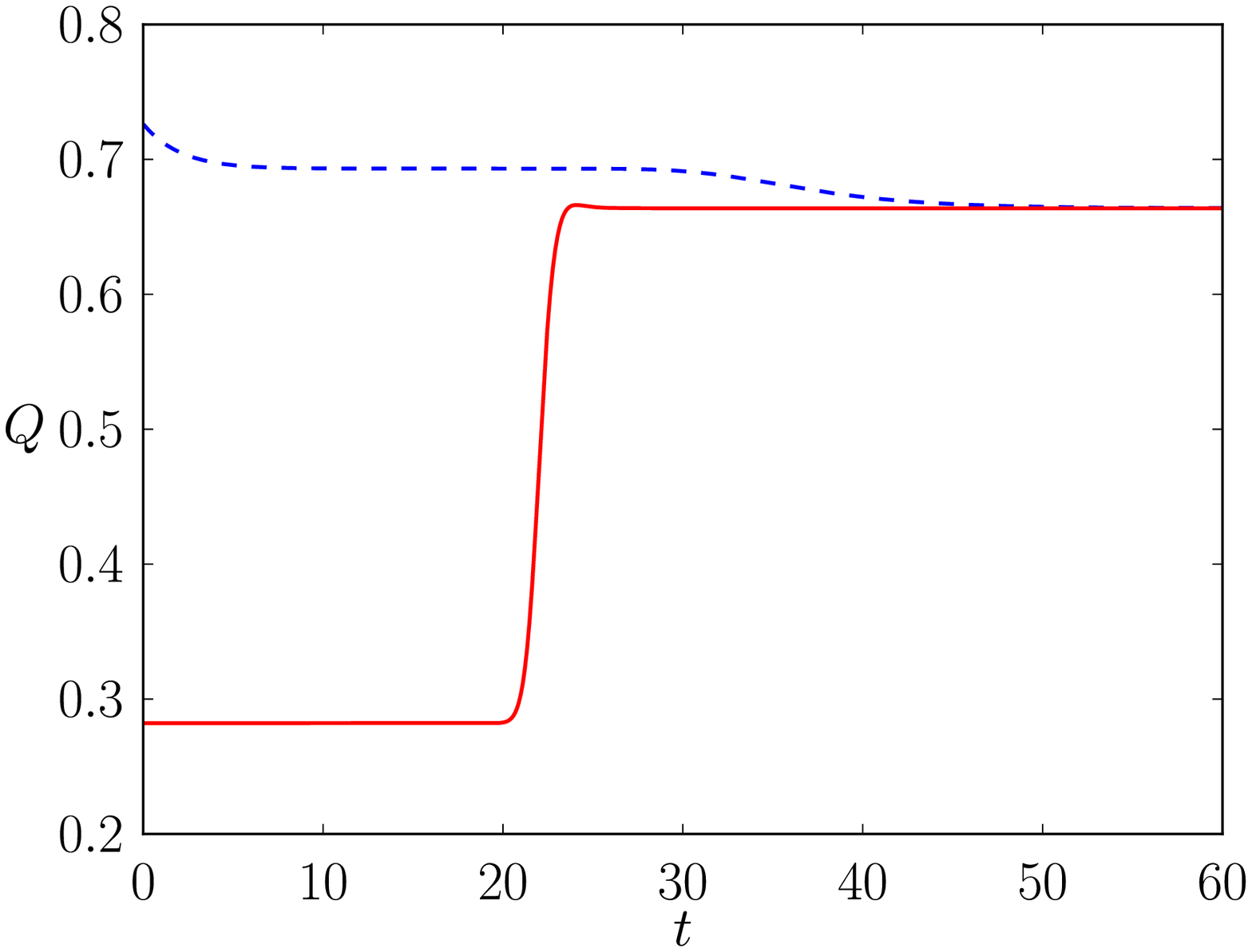}}
  \caption{\label{f:evoln_no_sup}
    Non-superradiant evolution
    ($M = 1$, $Q = 0.282$, $m = 0$, $q = 28.2$, $\omega = 10$, $\bar r_0 = 20$,
    $\bar \sigma = 1$, $A = 0.01$).
    Top: black hole mass $M_\mathrm{BH}$ (solid red), 
    Bondi mass $M_\mathrm{B}$ (dashed blue) and
    irreducible mass $M_\mathrm{irr}$ (dotted green).
    Bottom: horizon charge $Q(r_\mathrm{h})$ (solid red) and
    total charge at future null infinity $Q(r=1)$ (dashed blue).
  }
\end{figure}

For most of the evolutions shown here, we choose the scalar field mass to 
vanish, $m = 0$.
The scalar field charge is taken to be $q = 100/\sqrt{4\pi}=28.2$.
The parameters of the Reissner-Nordstr\"om background solution are 
$M=1$ and $Q=1/\sqrt{4\pi}=0.282$.
(The relatively large value of $q$ is chosen so that there is enough
  room for superradiance to occur, as the maximum superradiant frequency
  is proportional to $q$, see Eq.~\eqref{e:omegamax} below.)
For the initial data parameters in \eqref{e:iniphi} we first choose
$\omega = 10$, $A = 0.01$, $\bar \sigma = 1$, and $\bar r_0 = 20$. 
The corresponding evolution is shown in Fig.~\ref{f:evoln_no_sup}.
It shows the ``normal'' behavior we expect for a scalar field that mainly falls
into the black hole, while only a small amount escapes to infinity:
the black hole mass increases by about $13\%$ to almost the initial value of 
the Bondi mass, while the Bondi mass only decreases by a small amount;
similar behavior is seen in the evolution of the charges at the horizon and 
at $\scri$.
The energy radiated away at infinity $\Delta M_\mathrm{B}=0.021$
  is much smaller than the energy of the initial scalar field pulse 
  $E_\phi^\mathrm{initial}=0.15$.

\begin{figure}
  \centerline{\includegraphics[width=.475\textwidth]{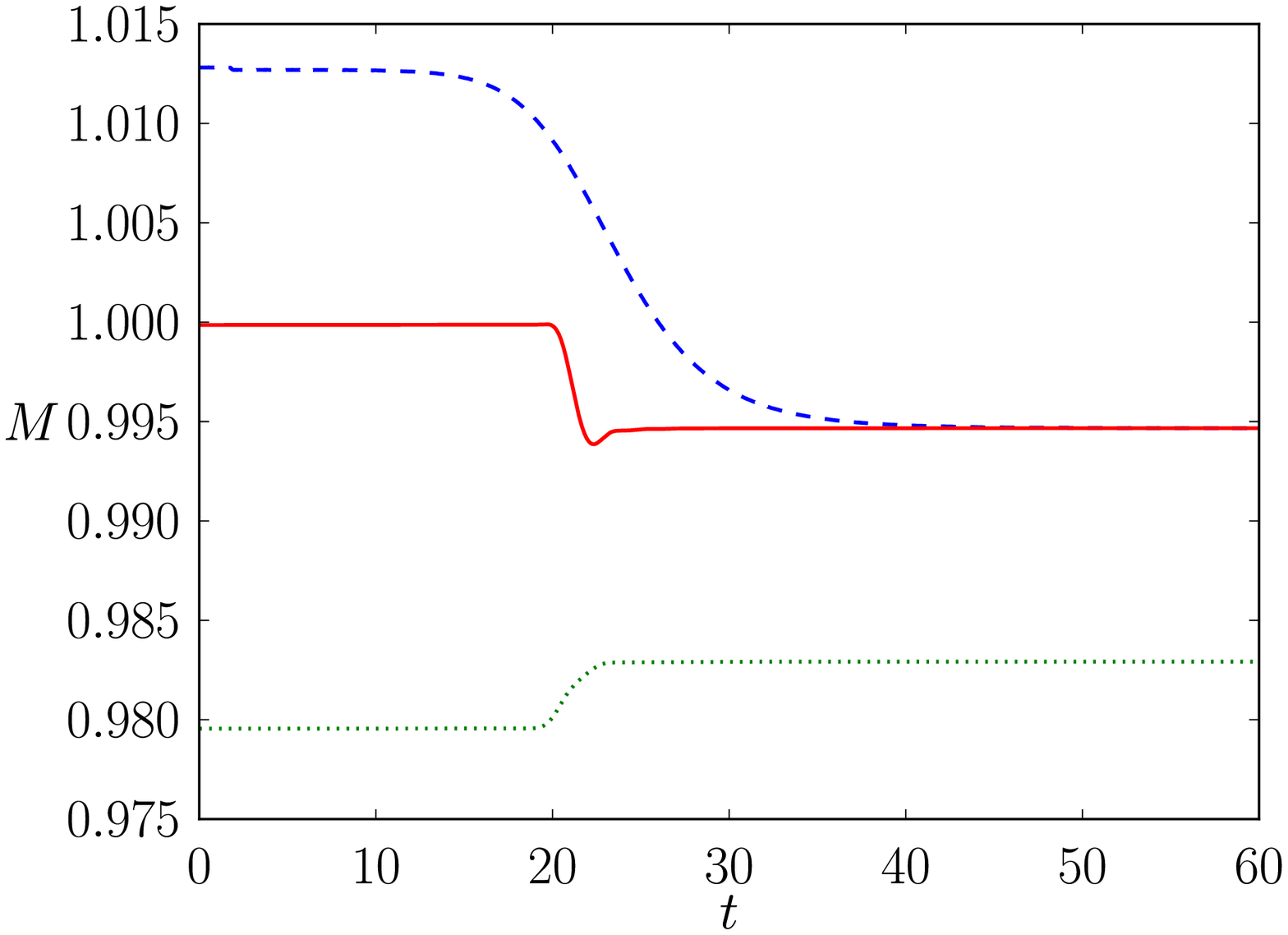}}
  \centerline{\includegraphics[width=.475\textwidth]{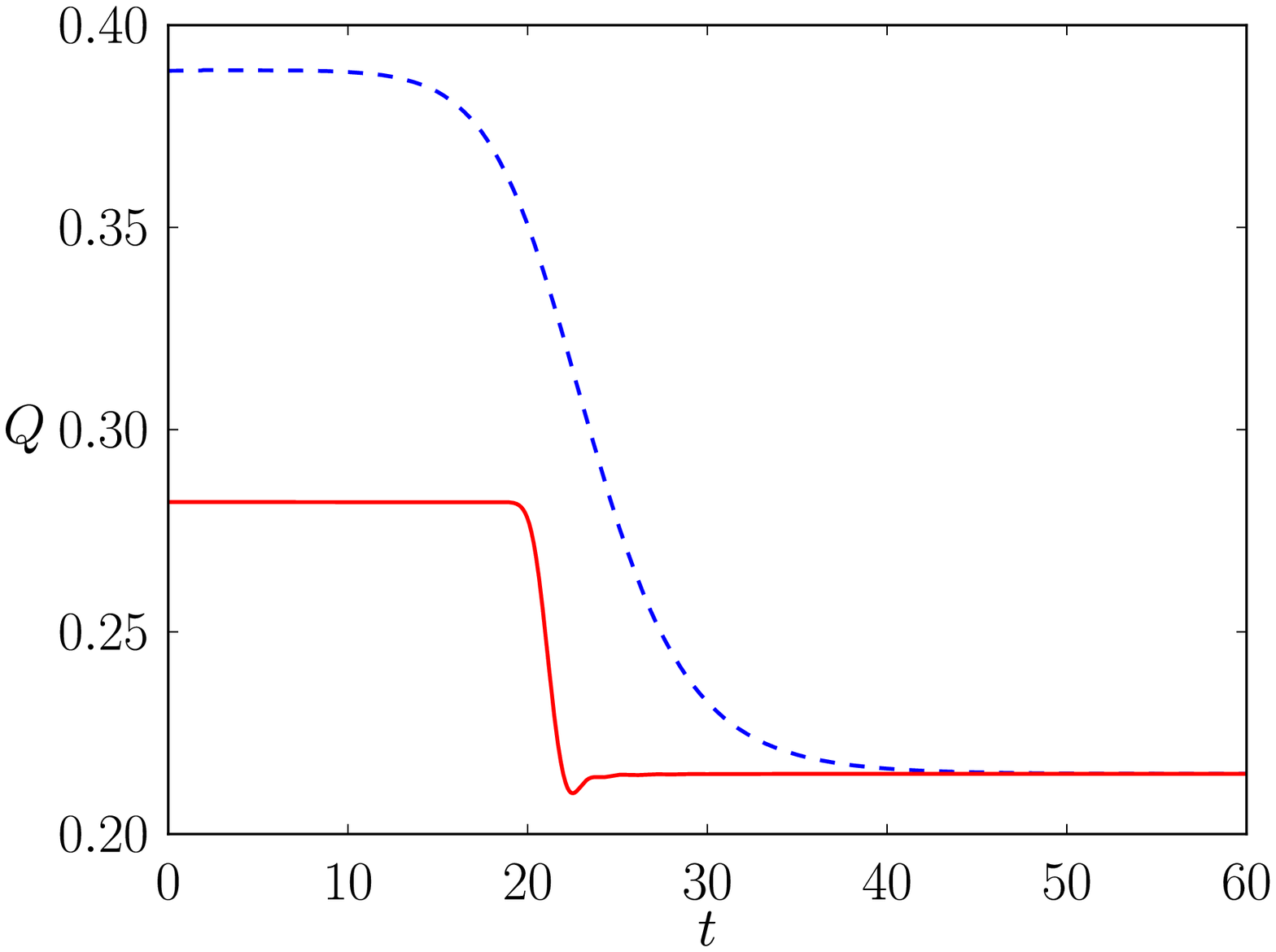}}
  \caption{\label{f:evoln_sup}
    Superradiant evolution
    ($M = 1$, $Q = 0.282$, $m = 0$, $q = 28.2$, $\omega = 2.4$, $\bar r_0 = 20$,
    $\bar \sigma = 1$, $A = 0.01$).
    Top: black hole mass $M_\mathrm{BH}$ (solid red), 
    Bondi mass $M_\mathrm{B}$ (dashed blue) and
    irreducible mass $M_\mathrm{irr}$ (dotted green).
    Bottom: horizon charge $Q(r_\mathrm{h})$ (solid red) and
    total charge at future null infinity $Q(r=1)$ (dashed blue).
  }
\end{figure}

Next, we change the frequency to $\omega = 2.4$, leaving all the other 
parameters unchanged.
The resulting evolution (Fig.~\ref{f:evoln_sup}) is markedly different:
now the black hole mass \emph{decreases} by about $0.5\%$ (and its final value
agrees with the final Bondi mass)---a clear indication of superradiance.
The horizon charge decreases by about $24\%$, the total charge at $\scri$ 
is nearly halved.
This time the energy radiated away at infinity
  $\Delta M_\mathrm{B}=0.018$ is larger than the energy of the initial scalar
  field pulse $E_\phi^\mathrm{initial}=0.01$.

\begin{figure}
\centerline{\includegraphics[width=.475\textwidth]{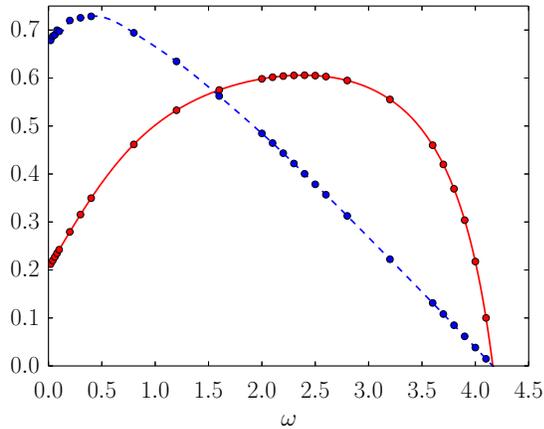}}
\caption{\label{f:efficiency}
  Superradiant efficiency $\eta$ (Eq.~\eqref{e:effi}, solid red) and 
  $\hat \eta$ (Eq.~\eqref{e:effi2}, dashed blue)
  as a function of $\omega$ for
  $M = 1$, $Q = 0.282$, $m = 0$, $q = 28.2$, $\bar r_0 = 20$, $\bar \sigma = 1$, 
  $A = 0.01$.
  Shown are results from several simulations (dots) and a cubic spline 
  interpolant.
}
\end{figure}

For this choice of parameters, the superradiant efficiency $\eta$ defined in
\eqref{e:effi} is shown as a function of $\omega$ in Fig.~\ref{f:efficiency}.
The evolution in Fig.~\ref{f:evoln_sup} corresponds to the maximum efficiency
of $\eta = 0.61$. 
This is comparable to the maximum efficiency reported by East \etal
\cite{East2014} for superradiant scattering of gravitational waves off a 
rotating black hole, $\eta \approx 0.58$.
There is a sharp decrease of $\eta$ towards zero at $\omega_\textrm{max} = 4.2$,
beyond which there is no superradiance. 

The alternative definition \eqref{e:effi2} of the superradiant
efficiency $\hat \eta$ is also plotted in Fig.~\ref{f:evoln_sup}.
Compared with $\eta$, the maximum of $\hat \eta$ is shifted towards
lower frequencies, with a peak value of $\approx 0.73$.
The cut-off frequency $\omega_\textrm{max}$ is virtually the same for
both definitions of the efficiency.
The quantity $\hat \eta$ shows an almost linear decrease as $\omega
\rightarrow \omega_\textrm{max}$.

Comparing our findings with perturbative results for monochromatic waves
  is not straightforward because we work with wave packets of finite extent, 
  the black hole mass and charge change drastically during the scattering 
  process,
and the amount of this change depends on the frequency.
  From monochromatic perturbation theory one would expect superradiance to
  vanish as $\omega \rightarrow 0$, and the upper cutoff frequency should be at \cite{Brito2015,DiMenza2015}
  \begin{equation}
    \label{e:omegamax}
    \omega_\mathrm{max} = \frac{qQ}{\bar r_+}.
  \end{equation}
For the initial black hole parameters in our simulation this results in
  $\omega_\mathrm{max} = 4.06$, quite close to the observed cutoff in the
  efficiency in Fig.~\ref{f:efficiency}.

\begin{figure}
  \centerline{\includegraphics[width=.475\textwidth]{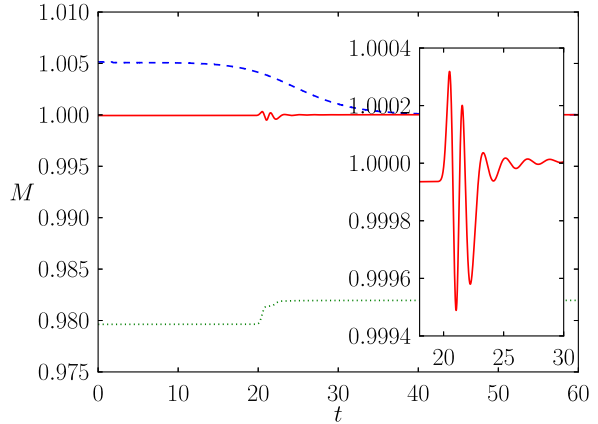}}
  \centerline{\includegraphics[width=.475\textwidth]{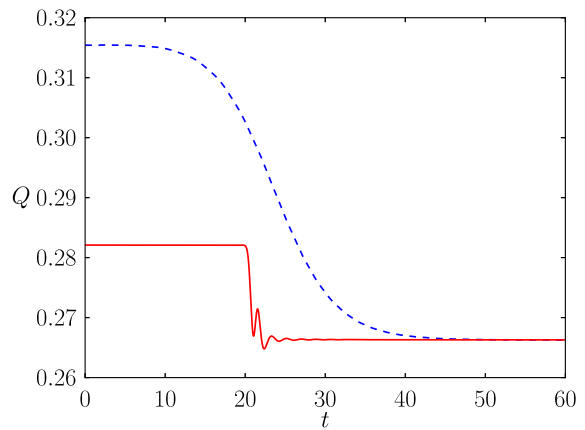}}
  \caption{\label{f:evoln_nonsup_Qdecr}
    Non-superradiant evolution with decreasing charge
    ($M = 1$, $Q = 0.282$, $m = 0$, $q = 28.2$, $\omega = 1.5$, $\bar r_0 = 20$,
    $\bar \sigma = 0.5$, $A = 0.01$).
    Top: black hole mass $M_\mathrm{BH}$ (solid red), 
    Bondi mass $M_\mathrm{B}$ (dashed blue) and
    irreducible mass $M_\mathrm{irr}$ (dotted green).
    Bottom: horizon charge $Q(r_\mathrm{h})$ (solid red) and
    total charge at future null infinity $Q(r=1)$ (dashed blue).
  }
\end{figure}

In Fig.~\ref{f:evoln_nonsup_Qdecr} we report an anomalous case where the black 
hole charge decreases even though the black hole mass increases by a small 
amount (about $0.1\%$) and hence there is no superradiance.
The scalar field charge $q$ is still positive in this evolution;
note however that the charge current density \eqref{e:31je} may still be 
negative depending on the form of the scalar field $\tilde \phi$, which is
indeed what we observe.

Similar behavior (decreasing charge, increasing mass of the black hole)
is found generically for negative $q$.

\begin{figure}
  \centerline{\includegraphics[width=.475\textwidth]{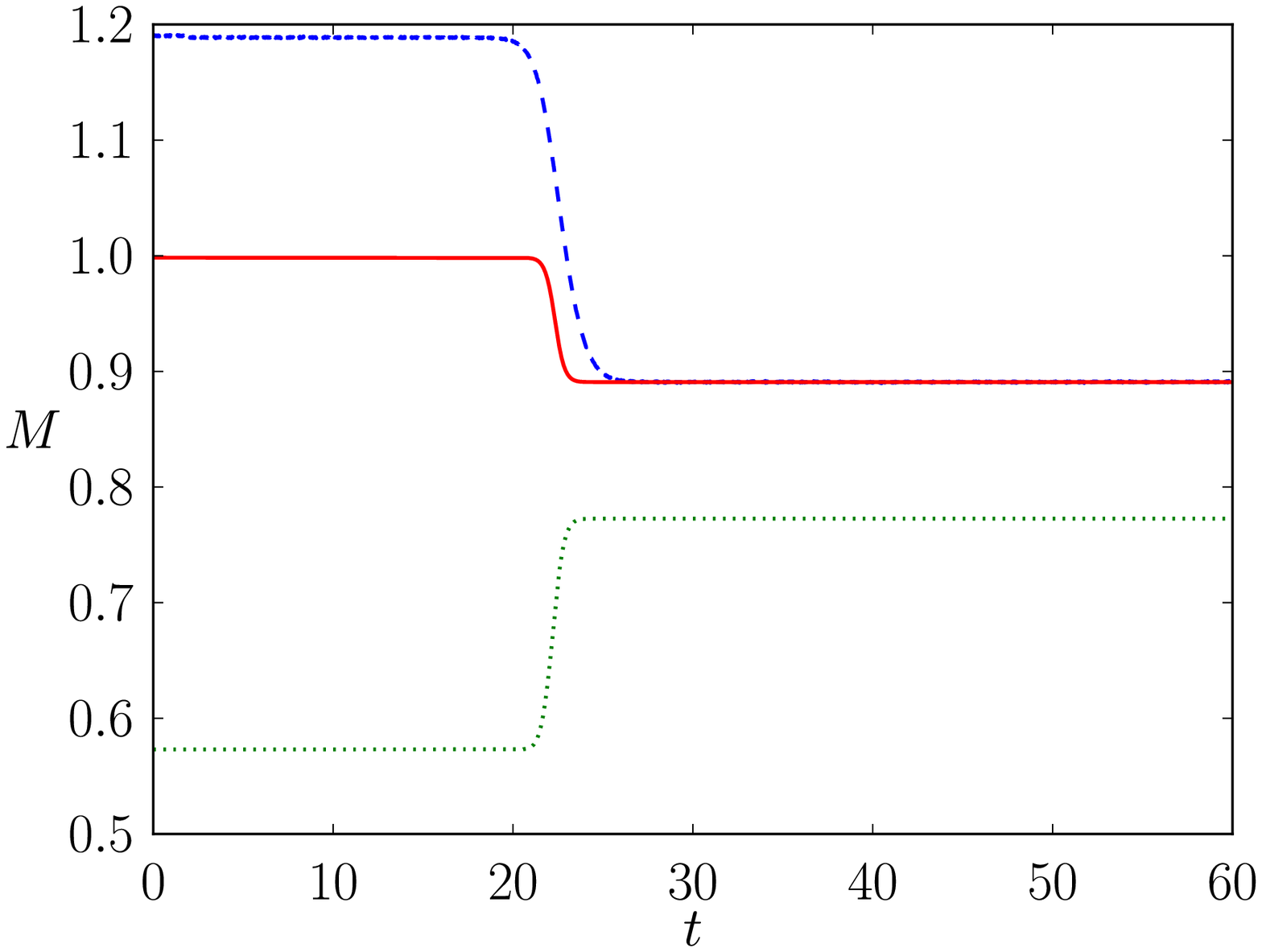}}
  \centerline{\includegraphics[width=.475\textwidth]{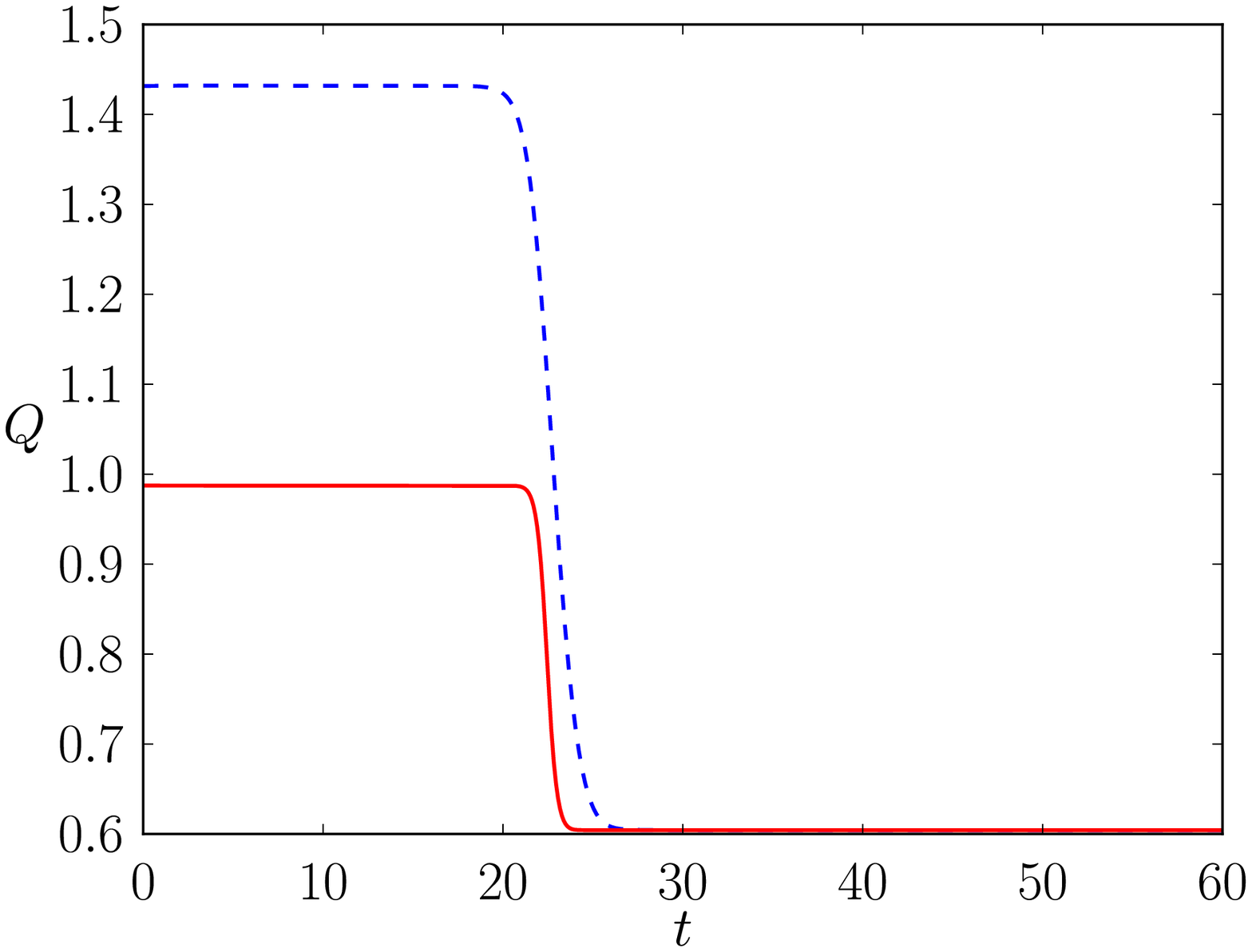}}
  \caption{\label{f:evoln_e}
    Superradiant evolution for a near-extreme black hole
    ($M = 1$, $Q = 0.987$, $m = 0$, $q = 28.2$, $\omega = 10$, $\bar r_0 = 20$,
    $\bar \sigma = 1$, $A = 0.01$).
    Top: black hole mass $M_\mathrm{BH}$ (solid red), 
    Bondi mass $M_\mathrm{B}$ (dashed blue) and
    irreducible mass $M_\mathrm{irr}$ (dotted green).
    Bottom: horizon charge $Q(r_\mathrm{h})$ (solid red) and
    total charge at future null infinity $Q(r=1)$ (dashed blue).
  }
\end{figure}

While so far the charge-to-mass ratio of the Reissner-Nordstr\"om black hole
has been moderate ($Q/M = 0.282$), Fig.~\ref{f:evoln_e} shows the evolution 
of a near-extreme black hole with $Q/M = 0.987$.
This case is superradiant with efficiency $\eta = 0.35$, $\hat \eta = 0.56$.

\begin{figure}
  \centerline{\includegraphics[width=.475\textwidth]{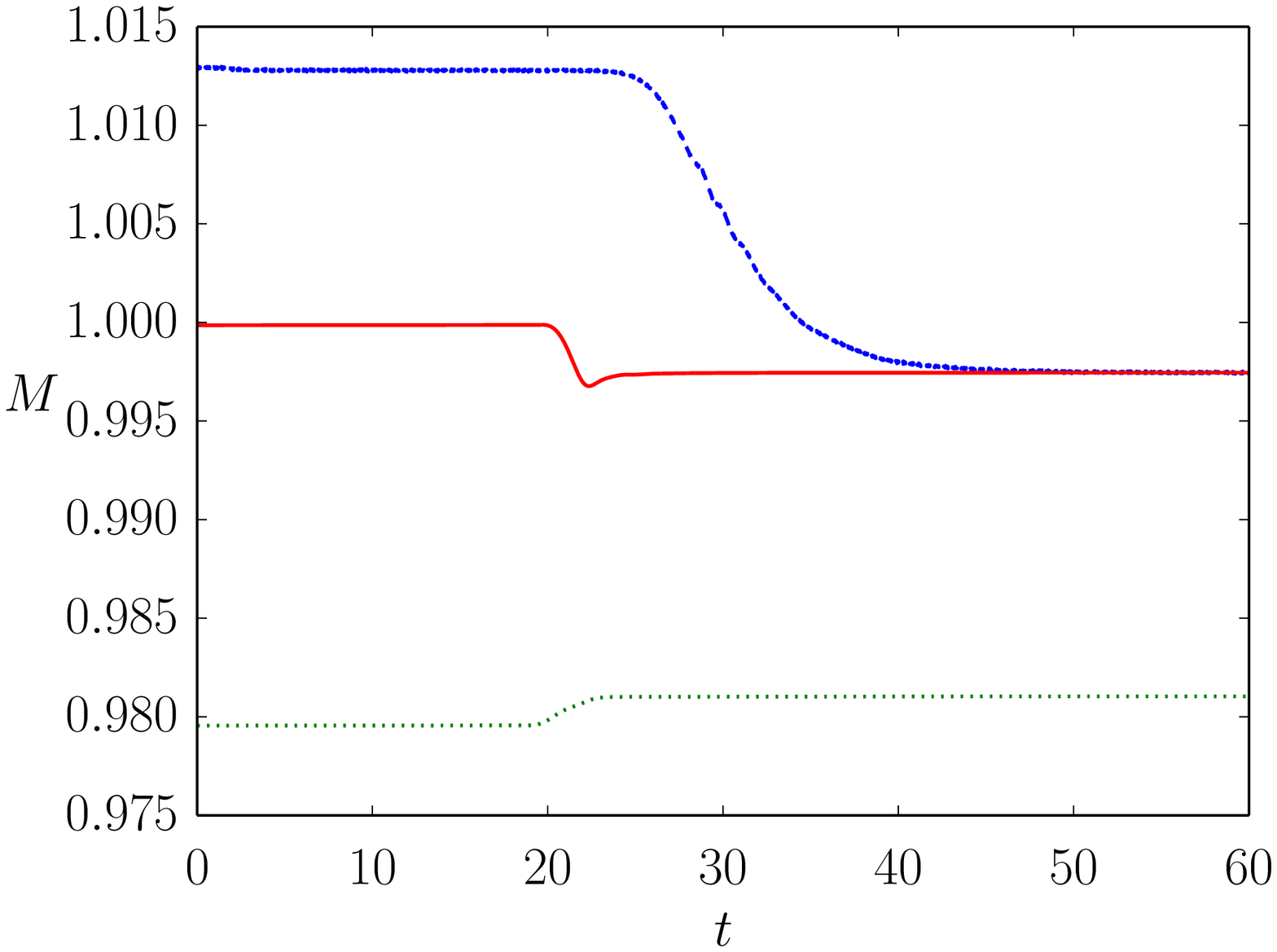}}
  \centerline{\includegraphics[width=.475\textwidth]{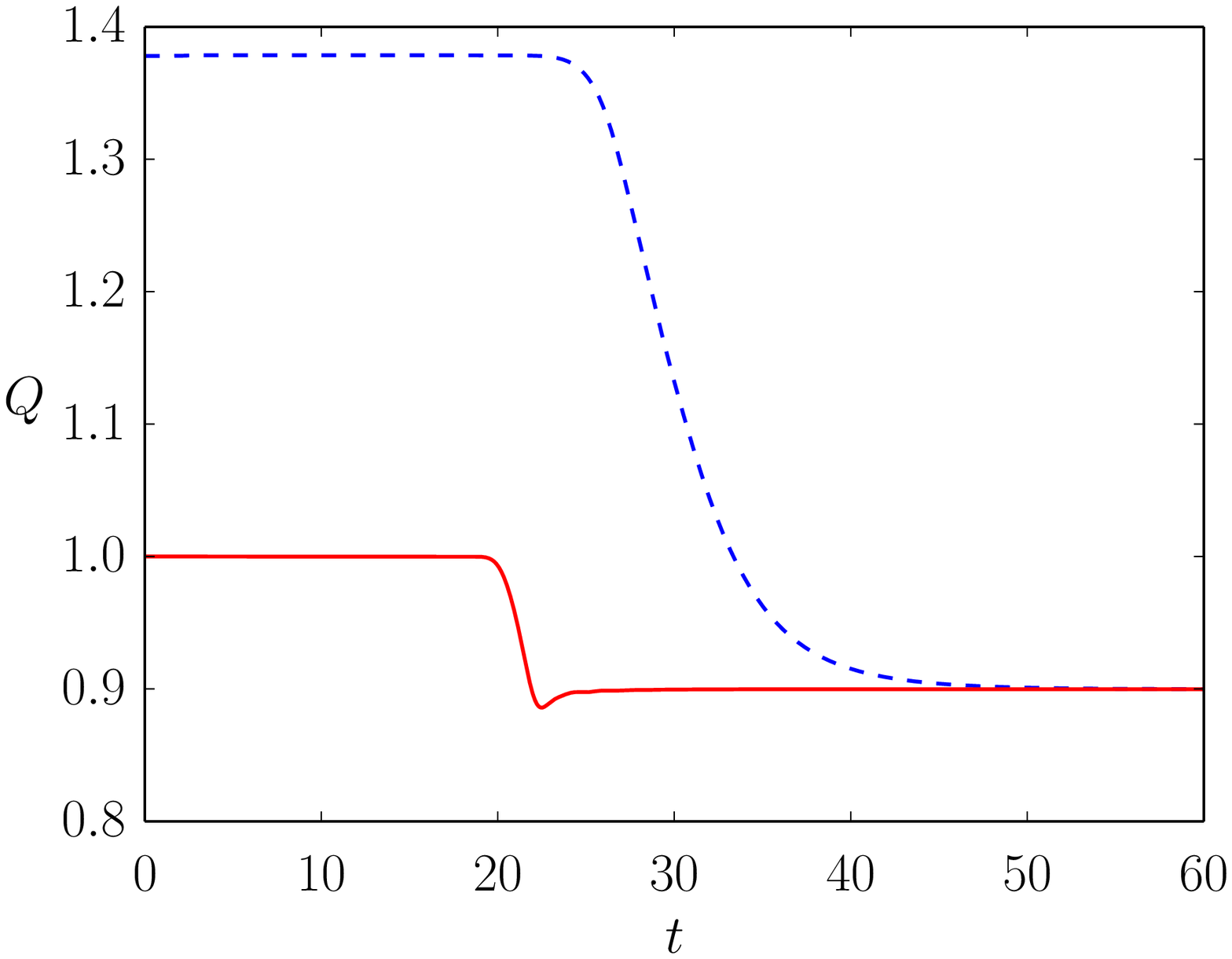}}
  \caption{\label{f:evoln_m0.5}
    Superradiant evolution with scalar field mass
    ($M = 1$, $Q = 0.282$, $m = 0.5$, $q = 28.2$, $\omega = 2.4$, 
    $\bar r_0 = 20$, $\bar \sigma = 1$, $A = 0.01$).
    Top: black hole mass $M_\mathrm{BH}$ (solid red), 
    Bondi mass $M_\mathrm{B}$ (dashed blue) and
    irreducible mass $M_\mathrm{irr}$ (dotted green).
    Bottom: horizon charge $Q(r_\mathrm{h})$ (solid red) and
    total charge at future null infinity $Q(r=1)$ (dashed blue).
  }
\end{figure}

\begin{figure}
  \centerline{\includegraphics[width=.475\textwidth]{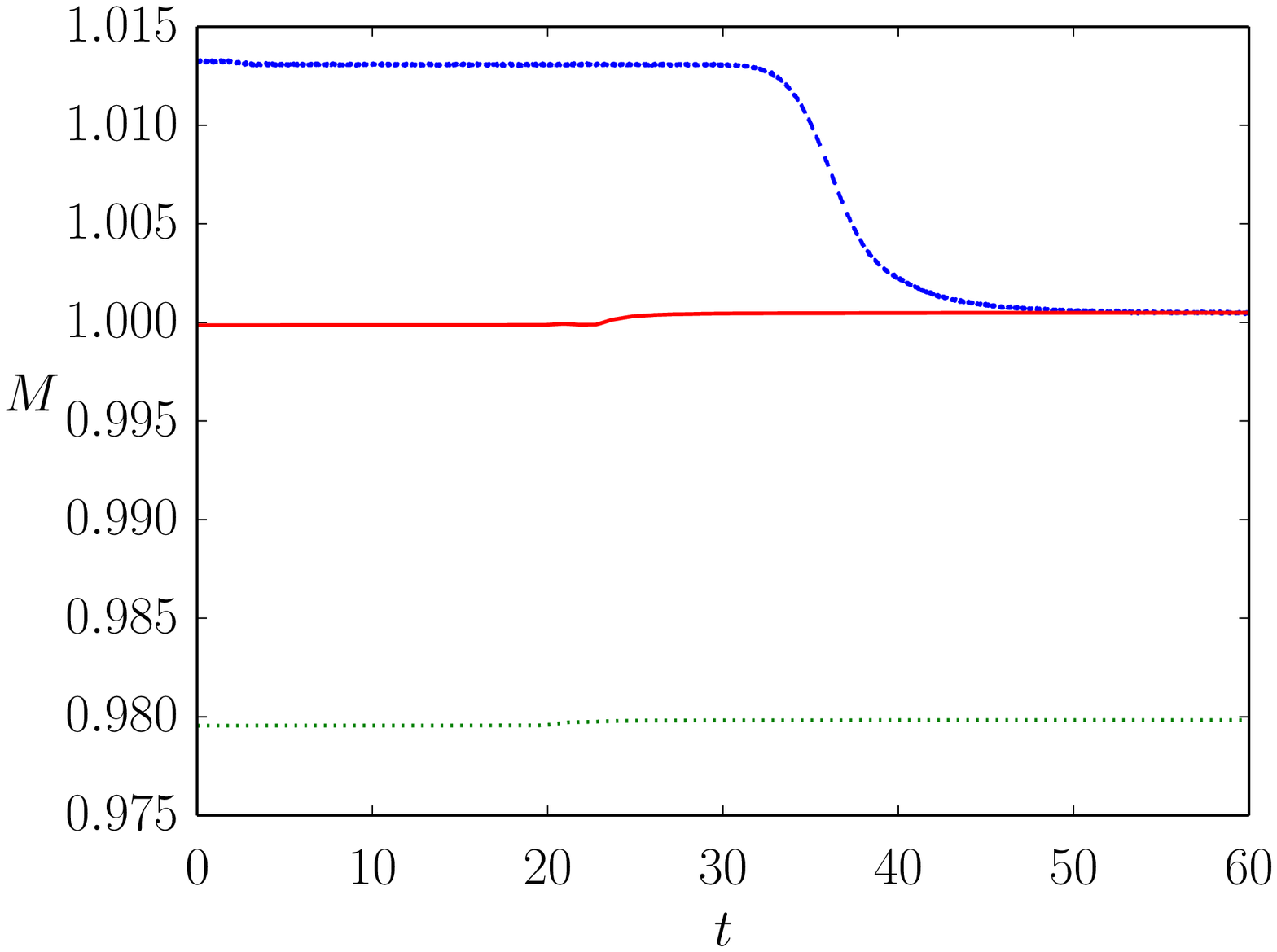}}
  \centerline{\includegraphics[width=.475\textwidth]{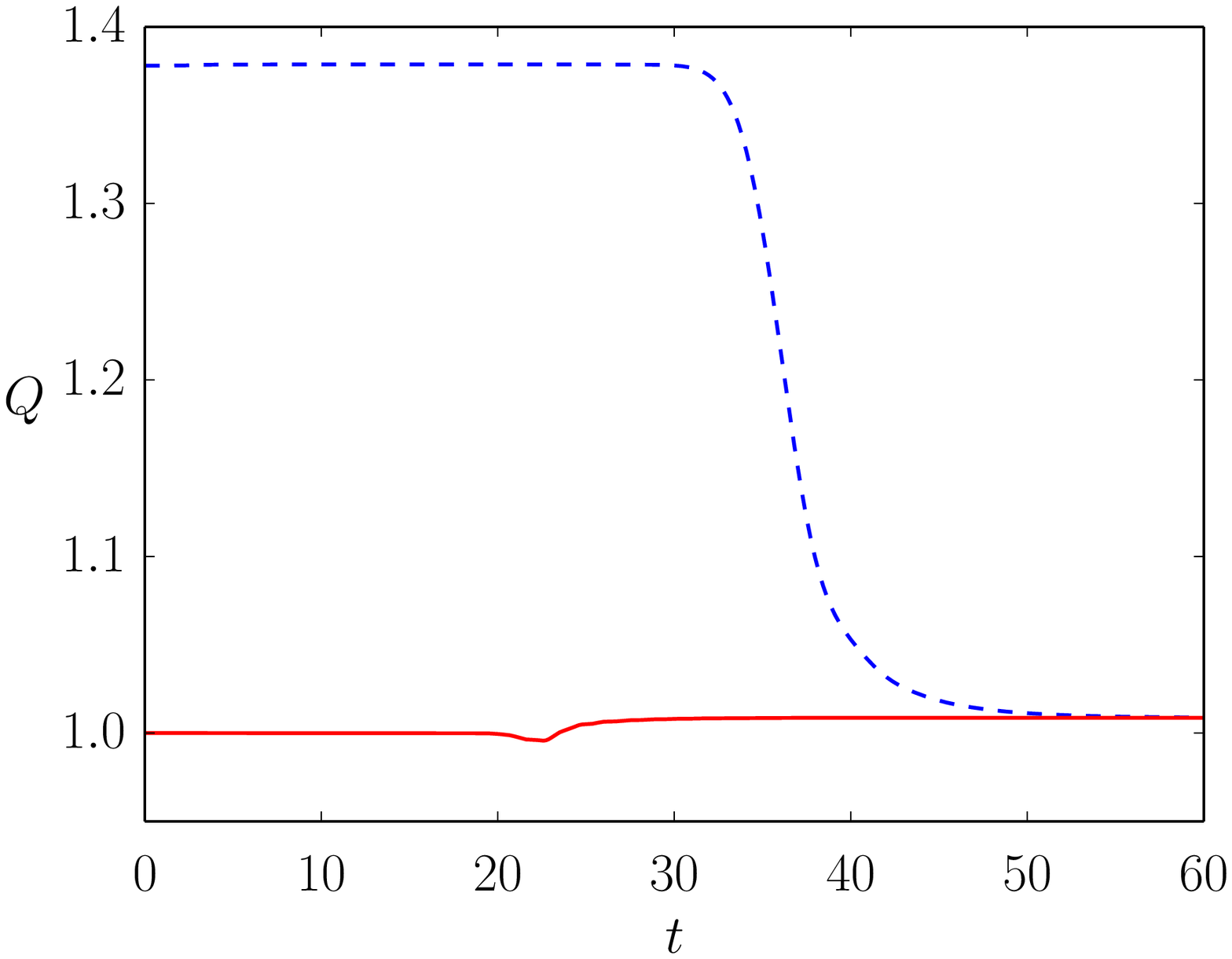}}
  \caption{\label{f:evoln_m1}
    Non-superradiant evolution with scalar field mass
    ($M = 1$, $Q = 0.282$, $m = 1$, $q = 28.2$, $\omega = 2.4$, 
    $\bar r_0 = 20$, $\bar \sigma = 1$, $A = 0.01$).
    Top: black hole mass $M_\mathrm{BH}$ (solid red), 
    Bondi mass $M_\mathrm{B}$ (dashed blue) and
    irreducible mass $M_\mathrm{irr}$ (dotted green).
    Bottom: horizon charge $Q(r_\mathrm{h})$ (solid red) and
    total charge at future null infinity $Q(r=1)$ (dashed blue).
  }
\end{figure}

Finally, we present evolutions with non-zero scalar field mass.
In Fig.~\ref{f:evoln_m0.5} we choose $m=0.5$; the other parameters are
the same as in the previous evolutions.
This case is superradiant with efficiency $\eta = 0.62$, $\hat \eta = 0.18$.
For $m=1$ (Fig.~\ref{f:evoln_m1}) we obtain a non-superradiant evolution.

As shown in \cite{DiMenza2015} in the test field case, there is no 
superradiance if $m \geqslant |qQ/\bar r_+|$, which evaluates to 
$m < 4.06$ for our choice of parameters.
Our results are consistent with this but note that the particular
value of $m$ at which superradiance ceases to exist depends on the
frequency $\omega$, and can be lower than this bound
(in our simulation already $m=1$ was non-superradiant).

For comparison with \cite{DiMenza2015} we also compute the (physical) radius
of the effective ergosphere (defined therein as the region where the 
potential becomes negative).
In the $m=0.5$ case this is $\bar r_\mathrm{ergo} = 16.9$,
in the $m=1$ case $\bar r_\mathrm{ergo} = 9$.
In both cases the initial data are concentrated around $\bar r_0 = 20$
with width $\bar \sigma = 1$ so the initial data lie outside the effective 
ergosphere. 


\subsection{Quasi-normal modes and tails}
\label{s:qnmtails}

\begin{figure}
\centerline{\includegraphics[width=.475\textwidth]{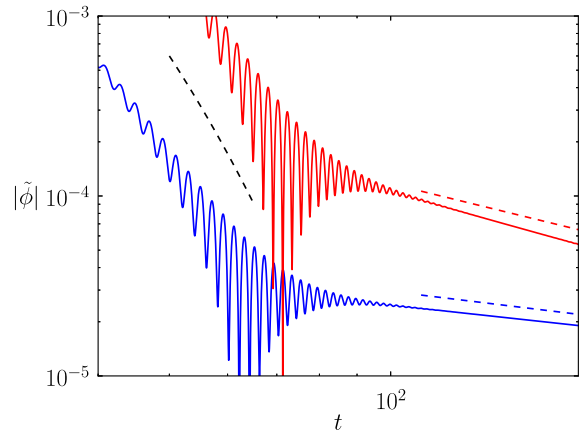}}
\caption{\label{f:decay}
  Absolute value of the scalar field at the horizon (upper red curve) 
  and at future null infinity (lower blue curve) as functions of time
  in a log-log plot.
  The predicted exponential decay during the quasi-normal
  mode phase (dashed line on the left) and the power-law decay during 
  the tail phase (dashed lines on the right) are shown for comparison.
  The same parameters as in Fig.~\ref{f:evoln_sup} are used
  ($M = 1$, $Q = 0.282$, $m = 0$, $q = 28.2$, $\omega = 2.4$, $\bar r_0 = 20$,
  $\bar \sigma = 1$, $A = 0.01$).
}
\end{figure}

At late times we observe a decay of the scalar field that is well described by 
an exponential ringdown followed by a power-law tail
(Fig.~\ref{f:decay}; here we return to the massless case $m=0$).

Quasi-normal modes of charged scalar fields on a 
Reissner-Nordstr\"om (and more generally Kerr-Newman) background
were computed in \cite{Hod2012,Konoplya2013}.
The relevant regime for the simulation shown in Fig.~\ref{f:decay} is that of 
large $qQ$, in which the authors find asymptotically
$\phi \sim \exp (-i\omega_\mathrm{QNM} t)$
with the fundamental ($n=0$) quasi-normal mode frequency 
\begin{equation}
  \label{e:predqnm}
  \omega_\mathrm{QNM} = \frac{qQ}{\bar r_+} 
  - i \frac{\bar r_+ - \bar r_-}{4 \bar r_+^2}.
\end{equation}
(In our case there is no rotation.)
The imaginary part of $\omega_\mathrm{QNM}$ (dashed line on the left in 
Fig.~\ref{f:decay}) agrees roughly with our simulation.
The real part predicted from \eqref{e:predqnm} is
$\mathrm{Re}\, \omega_\mathrm{QNM} = 3.07$
and the value measured from the simulation is $2.3$.
(Higher-order corrections in $1/(qQ)$ computed in \cite{Richartz2014} 
  are negligible in our case.)

A detailed analysis of late-time power-law tails of a charged scalar field on a 
Reissner-Nordstr\"om background was carried out by Hod and Piran in 
\cite{Hod1998}.
In the regime $|qQ| \gg 1$ their results imply for spherical perturbations
($\ell = 0$) that $\vert\phi\vert \sim t^{-1/2}$ at $\scri$ and 
$\vert\phi\vert \sim t^{-1}$ at the horizon. 
This agrees reasonably well with our simulation (Fig.~\ref{f:decay}).


\section{Conclusions}
\label{s:concl}

In this paper we investigated superradiance of a spherically symmetric 
charged scalar field scattering off a Reissner-Nordstr\"om black hole in 
asymptotically flat spacetime.
Unlike in previous studies, we solve the fully coupled 
Einstein-Maxwell-Klein-Gordon system.
A hyperboloidal evolution scheme on hypersurfaces of constant mean curvature
is used, which is ideally suited to black hole scattering experiments as these
slices extend smoothly from inside the horizon to future null infinity.

Our main result is that for sufficiently low frequency of the initial data, 
superradiance occurs and leads to substantial losses of mass and charge of
the black hole.
The maximum superradiant efficiency we observed, defined to be the change in 
black hole mass divided by the change in charge energy of the black hole,
was $\eta = 0.61$.
If the efficiency is defined instead by comparing the total Bondi mass
loss with the energy of the initial ingoing scalar field pulse, 
$\hat \eta = 0.73$ is obtained.
In the non-superradiant regime (increasing black hole mass) we found a 
somewhat anomalous case in which the black hole charge decreased even though the
scalar field charge was positive.
Our superradiant evolutions include near-extremal black holes ($Q/M = 0.987$) 
and nonzero scalar field mass ($m = 0.5$). 
It is clear that the massive case is challenging numerically due to
the terms $\sim \tilde \phi \, m^2/\Omega^2$ in the scalar field
evolution equation \eqref{e:eompsi}, which are formally singular at $\scri$ 
(and which vanish analytically only because the massive scalar field
falls off faster than any power of $\Omega$ at $\scri$ \cite{Winicour1988}).
While the evolutions presented here are stable, we do observe a
numerical instability for scalar field masses in the range 
$10^{-3} \lesssim m \lesssim 10^{-1}$ that we have not been able to cure.

We also analyzed the late-time decay of the massless scalar field and
found approximate
agreement with known perturbative results on quasi-normal modes and
power-law tails.

A nice by-product on the theoretical side is the construction of a conserved
current using the Kodama vector field, which we related to the Hawking mass 
as in \cite{Csizmadia2010}.
This enabled us to derive a Bondi mass loss formula.


\acknowledgments
We are grateful to Lars Andersson, Piotr Bizo\'n, Maciej Maliborski, 
Vincent Moncrief, and particularly Istv\'an R\'acz and Claudio Paganini 
for helpful discussions.
This research was supported by grant RI 2246/2 from the German Research 
Foundation (DFG) and a Heisenberg Fellowship to O.R.


\appendix

\section{Einstein evolution equations and regularity at 
  future null infinity}
\label{s:einsteinevoln}

Here we provide the (redundant) Einstein evolution equations that we do not 
enforce actively but monitor during the evolution.
The evolution equation for the conformal factor is
\begin{equation}
  \label{e:dtOmega}
  \dot \Omega = r X \Omega' - X \Omega + \tN(\half \Omega r^2 \Pi - \third K),
\end{equation}
and the traceless momentum obeys
\begin{eqnarray}
  \label{e:dtpi}
  \dot \Pi &=& r X \Pi' + 3 X \Pi + \tfrac{2}{3} r^{-1}(r^{-1}\tN')'\nonumber\\
  && + \tN \left[ -\tfrac{4}{3} \Omega^{-1} r^{-1} (r^{-1}\Omega')'
    - \tfrac{2}{3} \Omega^{-1} K \Pi  \right.\nonumber\\
  && \left. \qquad - \half r^2 \Pi^2 + 8\pi \Omega^2 r^{-2} 
    \tilde S^{\textrm{\scriptsize tr}\,rr}\right],
\end{eqnarray}
where
\begin{eqnarray}
  \label{e:31Strrr} 
  \tilde{S}^{\textrm{\scriptsize tr}\,rr}&=& \frac{2}{3} \left[ 
  |\tilde{\phi}'|^2 + iq \tilde{a}^r \left( \tilde{\phi}\tilde{\phi}'^* 
  - \tilde{\phi}^*\tilde{\phi}' \right)  
  \right. \nonumber\\
  && \quad + q^2 |\tilde{\phi}|^2 (\tilde{a}^r)^2 
  + \Omega^{-1} \Omega' \left( \tilde{\phi}^* \tilde{\phi}' 
  + \tilde{\phi} \tilde{\phi}'^* \right)  \nonumber\\
  &&\left. \quad + \Omega^{-2} |\tilde{\phi}|^2 \Omega'^2 
    - \frac{1}{4\pi}(\tilde{E}^r)^2 \right].
\end{eqnarray}
Though formally singular at $\scri$, equation \eqref{e:dtpi} is actually
regular provided the constraint equations hold (see \cite{Moncrief2009} for the general analysis).

In our case these conditions imply
\begin{eqnarray}
  \Omega &\hateq& 0, \label{e:scrO} \\
  \Omega' &\hateq& r\Omega'' \hateq -\frac{1}{3}K, \label{e:scrdO} \\
  \Omega''' &\hateq& - 8\pi |\tilde{\phi}|^2\Omega'^3, \label{e:scrdddO} \\
  \Pi &\hateq& 0, \label{e:scrpi} \\
  \Pi' &\hateq& -8\pi r^{-2}|\tilde{\phi}|^2\Omega'^2, \label{e:scrdpi}\\
  r \tilde{N}' &\hateq& \tilde{N}, \label{e:scrdN} \\
  \tilde{N}'' &\hateq& r^{-2}\tilde{N} 
  - 12\pi \tilde{N}|\tilde{\phi}|^2\Omega'^2, \label{e:scrddN} 
\end{eqnarray}
\begin{eqnarray}
  r X &\hateq& - \tilde{N}, \label{e:scrX} \\
  X' &\hateq& 0, \label{e:scrdX} \\
  X'' &\hateq& 12\pi r^{-1}\tilde{N}|\tilde{\phi}|^2\Omega'^2,
\end{eqnarray}
where $\hateq$ denotes equality at $\scri$.

If the scalar field mass is nonzero, $m \neq 0$, then the scalar field falls off
faster than any power of $\Omega$ towards $\scri$ \cite{Winicour1988},
which implies $\tilde \phi \hateq 0$.


\section{Alternative gauge conditions for the electromagnetic field}
\label{s:gaugefixing}
Here we present the different gauges we tried and explain why we choose 
\eqref{e:gauge} for our calculations. 
An obvious choice would be the physical Lorenz gauge $\nabla_\mu A^\mu = 0$. 
Rewriting this in terms of the conformal quantities yields
\begin{equation}
  \label{e:divA}
  \tilde{\nabla}_\mu \tilde{A}^\mu = 2 \Omega^{-1} \left[ \tilde{a}^i \Omega_{,i}
  + \frac{1}{3}\tilde{\Phi} \left( K - \Omega \tilde{K} \right) \right],
\end{equation}
which is manifestly singular at $\scri$. In particular, the evolution equation for $\tilde{\Phi}$ implied by this 
gauge condition as well as the scalar field evolution equation \eqref{e:eompsi} for $\tilde{\psi}$ would be singular.
Therefore we discard this gauge.

Another choice is the conformal Lorenz gauge 
$\tilde{\nabla}_\mu \tilde{A}^\mu=0$.
The evolution equation for $\tilde{\Phi}$ then becomes
\begin{eqnarray}
  \dot{\tilde{\Phi}} = \left( r X \tilde{\Phi} \right)' + 2 X \tilde{\Phi} - r^{-2} \left( r^2 \tN \tilde{a}^r \right)'.
\end{eqnarray}
Using the functions $f$ and $a$ defined in \eqref{e:cmcfunctions} and \eqref{e:cmcfunctions2}, 
the Reissner-Nordstr\"om background solution takes the form
\begin{equation}
  \tilde{\Phi}=\frac{Q f}{r B}, \qquad \tilde{a}^r=-\frac{Qa}{r B},
\end{equation}
where 
\begin{equation}
  B=1-\frac{2M}{r}+\frac{Q^2}{r^2}. 
\end{equation}
Unfortunately $a\rightarrow\infty$ and hence $\tilde a^r \rightarrow \infty$ 
at $\scri$, and so we discard this gauge condition as well.

A suitable gauge is $\tilde{\Phi}=0$, which leads to the condition
\begin{equation}
  \tilde{N} \tilde{\nabla}_\mu \tilde{A}^\mu = r^{-2} \left( r^2 \tilde{N} 
    \tilde{a}^r \right)'.
\end{equation}
This allows us to specify manifestly regular initial data, 
see Sec.~\ref{s:evolution}.


\bibliography{paper}

\end{document}